\documentclass[printer,traditabstract]{aa}

\usepackage{graphics,epsfig,amsmath,amssymb,amstext,txfonts}

\begin{document}
\title{Extended clustering analyses to constrain the deflection angular scale and source density of the UHECRs}
\author{Guillaume Decerprit$^1$ \and Nicolas G. Busca$^2$ \and Etienne Parizot$^2$}
\institute{$^1$DESY, Platanenallee 6, 15738 Zeuthen, Germany\\$^2$Laboratoire Astroparticule et Cosmologie (APC), Universit\'e Paris 7/CNRS, 10 rue A. Domon et L. Duquet, 75205 Paris Cedex 13, France.}

\offprints{guillaume.decerprit@desy.de}

\date{}

\abstract{The search of a clustering signal in the arrival directions of ultra-high-energy cosmic rays (UHECRs) is a standard method to assess the level of anisotropy of the data sets under investigation. Here, we first show how to quantify the sensitivity of a UHECR detector to the detection of anisotropy, and then propose a new method that pushes forward the study of the two-point auto-correlation function, enabling one to put astrophysically meaningful constraints on both the effective UHECR source density and the angular deflections that these charged particles suffer while they propagate through the galactic and intergalactic magnetic fields. We apply the method to simulated data sets obtained under various astrophysical conditions, and show how the input model parameters can be estimated through our analysis, introducing the notion of ``clustering similarity'' (between data sets), to which we give a precise statistical meaning. We also study how the constraining power of the method is influenced by the size of the data set under investigation, the minimum energy of the UHECRs to which it is applied, and a prior assumption about the underlying source distribution. We also show that this method is particularly adapted to data sets consisting of a few tens to a few hundreds of events, which corresponds to the current and near-future observational situation in the field of UHECRs.}


\authorrunning{G. Decerprit et al. }
\titlerunning{Constraining deflections and source density of UHECRs}
\maketitle

\section{Introduction}
\label{introduction}
Despite considerable experimental and theoretical efforts, the nature and origin of ultra-high-energy cosmic rays (UHECRs) remain largely unknown {(e.g. Kotera \& Olinto, 2011; Allard, 2011;  Abreu et al., 2011a,b; Tokuno et al., 2011; Sokolsky et al., 2010)}. It is believed that a detailed account of the processes responsible for the generation of what seems to be the highest energy particles in the universe would represent an important step towards the understanding of non-thermal cosmic sources in general, through the key process of particle acceleration in {astrophysical environments (see e.g. Allard \& Protheroe,~2009 and Hillas, 2006)}.

UHECRs are characterized by a very low flux, of the order of 1 per km$^{2}$ per century, and by a natural energy scale associated with their interaction with the intergalactic photons (notably from the cosmological microwave background): the GZK scale, around $6\,10^{19}$~eV. Above this energy, UHECR protons and nuclei lose energy through photo-pion production and photodissociation, respectively, so they cannot propagate over large distances. This results in the appearance of a ``GZK'' horizon beyond which UHECRs cannot be visible from Earth. At $6\,10^{19}$~eV, the horizon distance is of the order of 200 Mpc, and it is rapidly decreasing above this energy. As a consequence, the number of sources effectively observable, and thus the UHECR flux actually observed is predicted to decrease sharply above this energy (see Greisen, 1966 and Zatsepin, 1966), in agreement with the observations of HiRes~(Sokolsky et al., 2009) and Auger~(Abraham et al., 2010; Abreu P. et al., 2011a).

While it makes the observation of hundreds of UHECRs a very difficult experimental challenge~(Kajino et al., 2010), this so-called ``GZK effect'' can actually be regarded as a helping hand offered by Nature for the identification of UHECR sources. Indeed, it comes to alleviate the problem caused by the deflection of cosmic rays by the intervening magnetic fields (Galactic and extragalactic), as its main consequence is to reduce the number of contributing sources, hence to increase their angular separation over the sky. Now, if the deflections are smaller than the typical angular distance between the main UHECR sources (which is also more likely to happen at the highest energies, where the magnetic rigidity is larger), localized clusters of events can be expected to be identified as soon as the integrated exposure of the UHECR experiments becomes large enough for several events to be contributed by single sources.

This suggests that the analysis of UHECR clustering over the sky can be used to constrain the number of sources, or equivalently the source density, together with the effective deflection of UHECRs from their sources to the Earth. These two parameters are particularly important to understand the phenomenology of UHECRs, and eventually identify their sources. On one hand, the typical angular size of the deflections provides clues about the charge of the UHECR particles and/or the strength of the intervening magnetic fields. On the other hand, narrowing the range of possible source densities would directly lead to the rejection of some source models, notably if the number density of a given type of sources in the nearby universe is smaller than actually required by the data. In addition, an estimate of the UHECR source density would immediately reflect into an estimate of the individual source power, since the total energy input in the form of UHECRs is known from the measured flux. In turn, knowing the individual source power would provide important clues towards the possible sources (which must obviously be at least as powerful) as well as the acceleration efficiency (i.e. the fraction of energy imparted to cosmic rays), if the total source power is known or estimated by other means.

Observations of the clustering signal in the UHECR data from AGASA triggered several studies essentially designed to constrain the local density of sources (e.g. see~Dubovsky et al., 2000) along with constraints on the source luminosity~(Blasi \& De Marco, 2004). The clustering measurement brought by the Pierre Auger Observatory~(Abreu et al., 2010; Abraham et al., 2007) put on more solid grounds the power of such studies, and the analysis of the first Auger data lead to stronger constraints on the density of sources (see~Cuoco et al., 2008 and Abreu P. et al., 2011b) by using the autocorrelation function and performing a scan in energies and separation angles. An update of this result was recently presented (see~De~Domenico et al. in Abreu P. et al., 2011b), which notably constrained the density of sources assumed to be distributed like the galaxies.
In this context, our study adds a key parameter of the phenomenology of UHECRs and put an emphasis on the joint constraints that can be put on the source density and the deflection.

{Recently, the Auger Collaboration rejected the hypothesis that the UHECR sky was isotropic at a 99\% Confidence Level~(Abraham et al., 2007): despite its limited Confidence Level, this result represents an expected and important milestone in the field.} It was obtained by demonstrating that the arrival directions of UHECRs have a larger probability to lie within a given subpart of the sky (originally built as the region within 3 degrees of an AGN reported in a given catalog) than if they were isotropic. While this provided no clue about the actual UHECR sources, nor about their actual deflections by magnetic fields, it did show that the deflections are not large enough to spoil, in the \emph{observed} sky, what may be inferred as an anisotropic \emph{source} distribution. Likewise, the two-point correlation function of the Auger events reveals an excess of clustering at some angular scale, compared to isotropic expectations, with an overall significance of the order of 1\%~(Abraham et al., 2008). {An update of this result was released in 2010 (see~Abreu et al., 2010).}

In this paper, we investigate whether this kind of clustering analyses can be developed to do more than testing the isotropy hypothesis or setting limits on the sole density of sources, and actually derive constraints within the 2D parameter space that is most relevant from the astrophysical point of view, namely the space of UHECR source density and angular deflections, which we call here in short the ($n_{\mathrm{s}}$,$\delta$)-space.

\section{Principle of the analysis}
The idea behind the proposed study is that the ``level of clustering'' of a given data set (a notion to be understood in an intuitive way for the moment) depends on the number of sources contributing, and on the amplitude of the random deflections which spread the UHECRs around a central position, resulting in less dense clusters and an possible overlap of nearby sources. Qualitatively, if the local universe (within the GZK horizon) contains only a few sources, and the magnetic deflections are small, the data set will show a high level of clustering on small angular scale~(Mollerach et al., 2009). On the other hand, if there are many sources all over the sky and large deflections, data sets should not show a significant level of clustering at any scale. An intermediate situation would be that of a low source density, with only a few sources within the GZK horizon, but relatively large deflections. In that case, one may also expect some significant clustering, although at a larger angular scale. 

The analysis proposed here consists in comparing the level of clustering of a data set of UHECRs with the level of clustering of simulated data sets of the same size that would be produced by a range of astrophysical models characterized by a given source density, $n_{\mathrm{s}}$, and a given deflection scale, $\delta$. Models that have a very low probability to provide either ``as much clustering'' or ``as little clustering'' as the data set under investigation will then be declared incompatible with the data set (with the corresponding confidence level).

This analysis, however, depends on an assumption regarding the global source distribution underlying the simulated universes. Indeed, the level of clustering of a data set may also reflect the level of clustering of the sources themselves. For instance, a large density of sources distributed uniformly over the sky will produce a data set with no significant clustering, whatever the deflection scale, while a data set built from sources distributed non-homogeneously will be ``more clustered'' than a random, uniform data set, because the \emph{sources} are. For this reason, we investigate here two different cases, drawing sources randomly (with the density under investigation): i) from a perfectly uniform distribution (sources are equally likely to be found anywhere in the Universe), ii) from a distribution matching that of the galaxies in the nearby universe (in a similar way as in~Abreu P. et al., 2011b). The former case allows us to explore the main characteristics of this kind of analyses, in a general, ``neutral'' situation. The latter case is more relevant to the current UHECR phenomenology favoring bottom-up scenarios, where the sources are presumably distributed like matter itself, whether they are a class of galaxies (e.g. active galaxies, classes of AGNs...), a roughly random subset of galaxies (e.g. gamma-ray burst scenarios, young pulsars...), or a subset of clusters of galaxies (e.g. cluster-scale shocks...)

\section{Simplifying assumptions}
\label{sec:assumptions}
The deflection of a charged UHECR from its source to the Earth is governed by its interaction with the intervening magnetic field, both Galactic and extragalactic. Particles with the same magnetic \emph{rigidity}, i.e. the same energy-to-charge ratio, $E/Z$, follow identical paths in the universe, while the overall deflections are essentially proportional to $Z$ and inversely proportional to $E$. UHECRs from different sources in the sky should be deflected in different ways, because they folllow different paths in the Galaxy.

However, the amplitude and geometrical structure (orientation, coherence length, filling factor) of the magnetic fields in the universe are unfortunately not known at the moment, even grossly~(Beck, 2011). Therefore, in this exploratory paper, we do not attempt a detailed treatment of the UHECR deflections, which would be strongly model-dependent anyway, and consider a simplified model in which the deflections are represented by a single parameter, $\delta$, which represents the ``angular size'' of the overall deflection, for any cosmic ray above a given energy. We ascribe to this angular parameter $\delta$ the following operational meaning: the event directions are drawn from a 2D gaussian distribution centered at the position of the source and a spread such that the average deflexion angle is $\delta$. 

Note that, in reality, the distribution of UHECRs from a given source are expected to be somewhat elongated along the direction of the average regular magnetic field towards the source, with the highest rigidity particles being deflected the least. However, since the UHECR spectrum is steep, most of the UHECRs received form a given source above a given energy have essentially that energy and should thus experience similar deflections (except if they have different charges). For the same reason, we shall use a unique angular parameter $\delta$ for all events, whatever their energy above a given threshold (although $\delta$ does, of course, depend on that threshold). It would of course be easy to scale the deflection with the inverse of the rigidity, but that would not change our results significantly here: what we are interested in is a constraint on an effective ``deflection scale'', averaged over all UHECRs above a given energy, from all regions of the sky. More refined models can easily be adapted for the needs of future analyses with specific data sets and specific magnetic field and composition models. It may also be objected that, while the random (turbulent) magnetic field may be expected to essentially ``broaden'' the source into a roughly Gaussian distribution of UHECRs over the scale $\delta$, the regular magnetic field, mostly in the Galaxy, tends to shift the position of the centroid of this distribution away from the actual source position. However, since we shall draw the positions of the sources randomly (with a given source density and according to a given probability function), the actual position of the centroids of the resulting distributions of events are not relevant (unless the shifts are larger than the typical autocorrelation scale of the source probability distribution), and we can simply ignore such shifts for the global clustering analysis proposed here.

Finally, we assume that the deflections are independent of the source distance, which essentially comes down to assuming that they are dominated by the influence of the Galactic magnetic field, which appears reasonable given the current understanding of the extragalactic magnetic field (e.g.~Ensslin, 2005). However, just as with the above simplifications, it is easy, in principle, to include any specific law for the deflection of UHECRs from any given source when building our test data sets, at the cost of additional free parameters or model assumptions (which we believe unnecessary in the current stage of understanding of the UHECR phenomenology).

Another assumption concerns the luminosity of the different sources. Here, we simply assume that all sources have the same intrinsic luminosity. This may apply if the main sources are ``candle events'' (such as, possibly, GRBs~(Murase et al., 2008)), but it is almost certainly not the case if AGNs are the main sources. However, in the absence of a favored model for the UHECR sources, we chose not to add another set of free parameters to describe the luminosity distribution. The source density that the proposed analysis may favor with a real data set should then be interpreted as an effective source density, that is the density which leads to a similar ``clustering signal'' if all sources were intrinsically similar. This value remains an interesting parameter to constrain UHECR source models. We simply note here that data sets built with sources having a range of luminosities generically provide a larger ``clustering signal'' than data sets from identical sources with the same cosmic distribution and density. This is because the flux is then likely to be dominated by a source which is unusually close and intense, providing (on average) a larger number of UHECRs than in the case of perfectly identical sources.

\section{Data sets and clustering signal}
\label{sec:ctheta}

Our goal is to compare the level of clustering of a reference data set produced by a given astrophysical model, with the level of clustering of some tested data sets. A given astrophysical model is characterized here by a set of assumptions concerning:
\begin{itemize}
\item the probability distribution of the source locations: either uniform or proportional to the matter density in the universe;
\item the source density, $n_{\mathrm{s}}$: we explore densities ranging from $10^{-6}$ to $10^{-3}$ per Mpc$^{3}$, the lower limit being set by the requirement that there are at least a few sources{ within the GZK horizon\footnote{Due to energy losses of the protons, the sources at a greater distance than $\sim$200 Mpc can not contribute to more than a few percent to the UHECR flux above 60 EeV. This effect is the so-called ``GZK horizon'' and allows us to define a maximum distance for the sources.},} and the upper limit by the observation that there are no visible sources in our Galaxy or local group;
\item the angular scale, $\delta$, of the deflections -- which we leave as a free parameter ranging from 1 degree (the typical angular resolution of UHECR experiments) to 90 degrees.
\end{itemize}

For each model, we simulate data sets with a given number of events, $N_{\mathrm{evt}}$, in the following way. First, we draw the position of the sources in the universe as a random realization of the underlying source distribution, with source density $n_{\mathrm{s}}$. Then, we propagate the cosmic rays injected by these sources in the intergalactic medium with a power-law spectrum in $E^{-x}$, as required to fit the UHECR data. Here, for definiteness, we assume pure protons sources with a maximum energy $E_{\max} = 10^{20.7}$~eV, but we found that the results are essentially independent of $x$ and $E_{\max}$. We set a minimum energy $E_{\min}$ and draw $N_{\mathrm{evt}}$ events above that energy from the propagated flux at Earth, taking into account the detector acceptance. Here, we assume the same coverage map as that of the Pierre Auger Observatory. Finally, for each of the $N_{\mathrm{evt}}$ events, we choose its location at an angular distance $\alpha$ from its actual source, where $\alpha$ is drawn randomly from a 2D-Gaussian distribution with variance $\delta^{2}$ (see Sect.~\ref{sec:assumptions}).

\begin{figure}[!t]
\begin{center}
\includegraphics[width=0.83\linewidth]{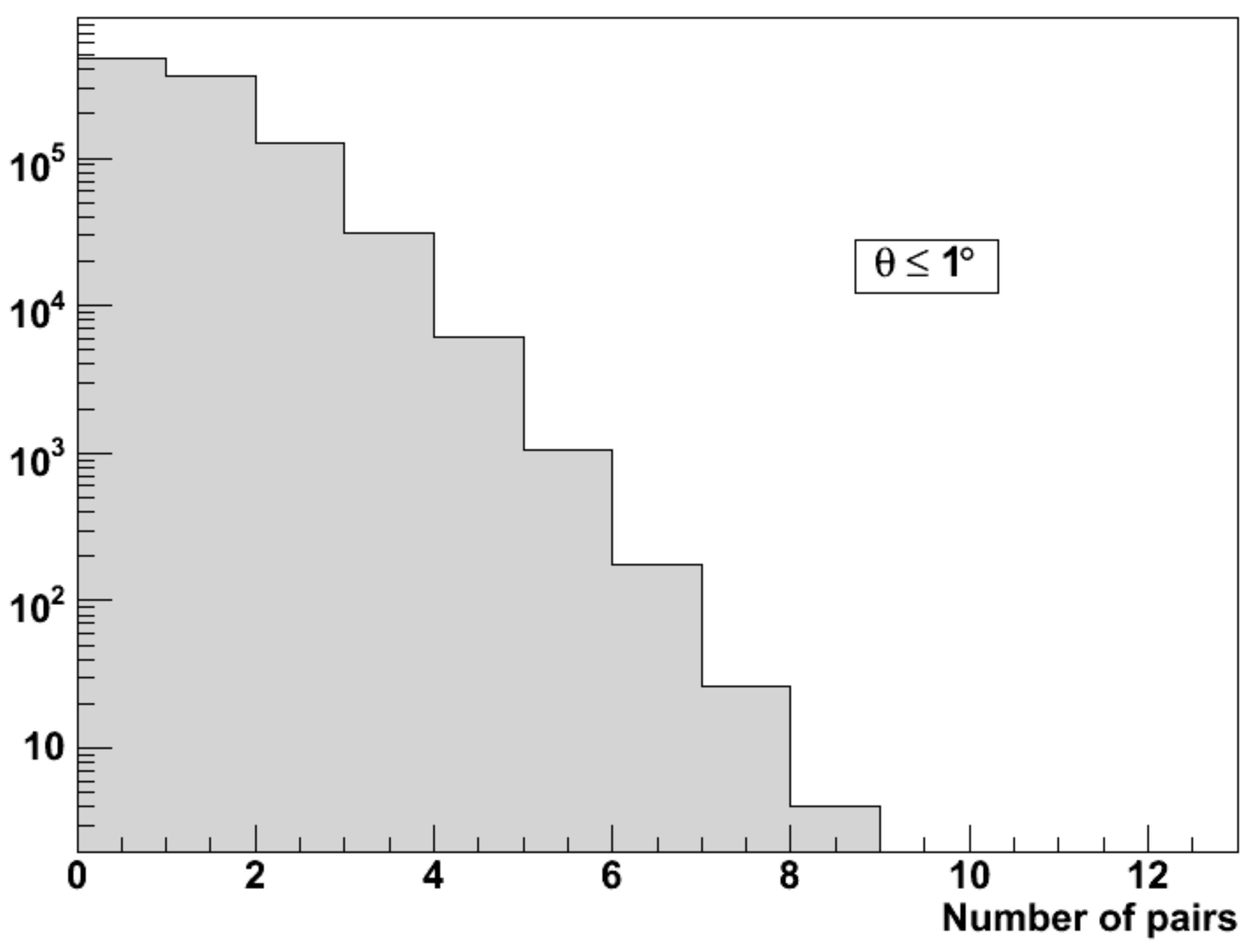} \\
\includegraphics[width=0.83\linewidth]{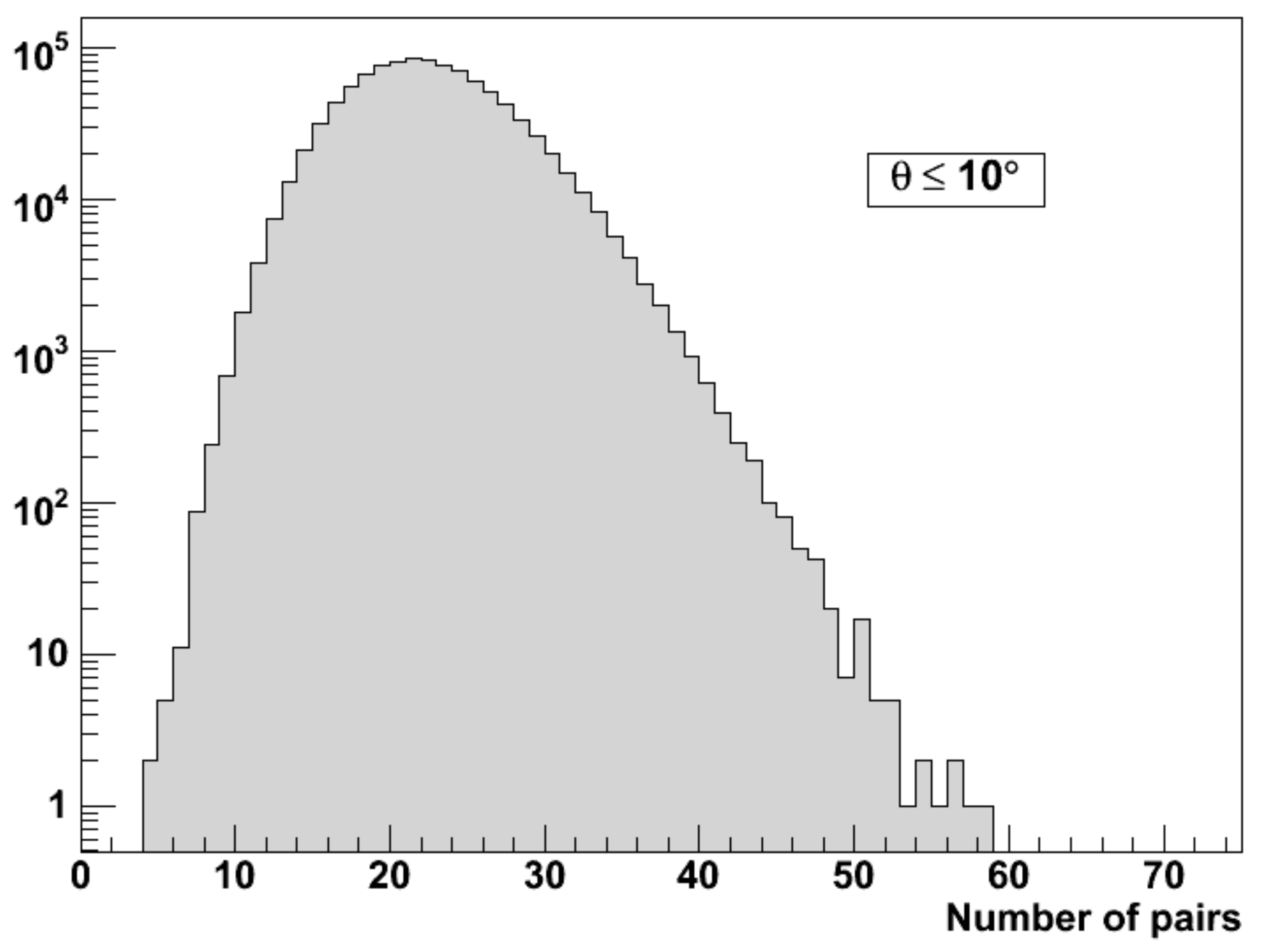}\\
\includegraphics[width=0.83\linewidth]{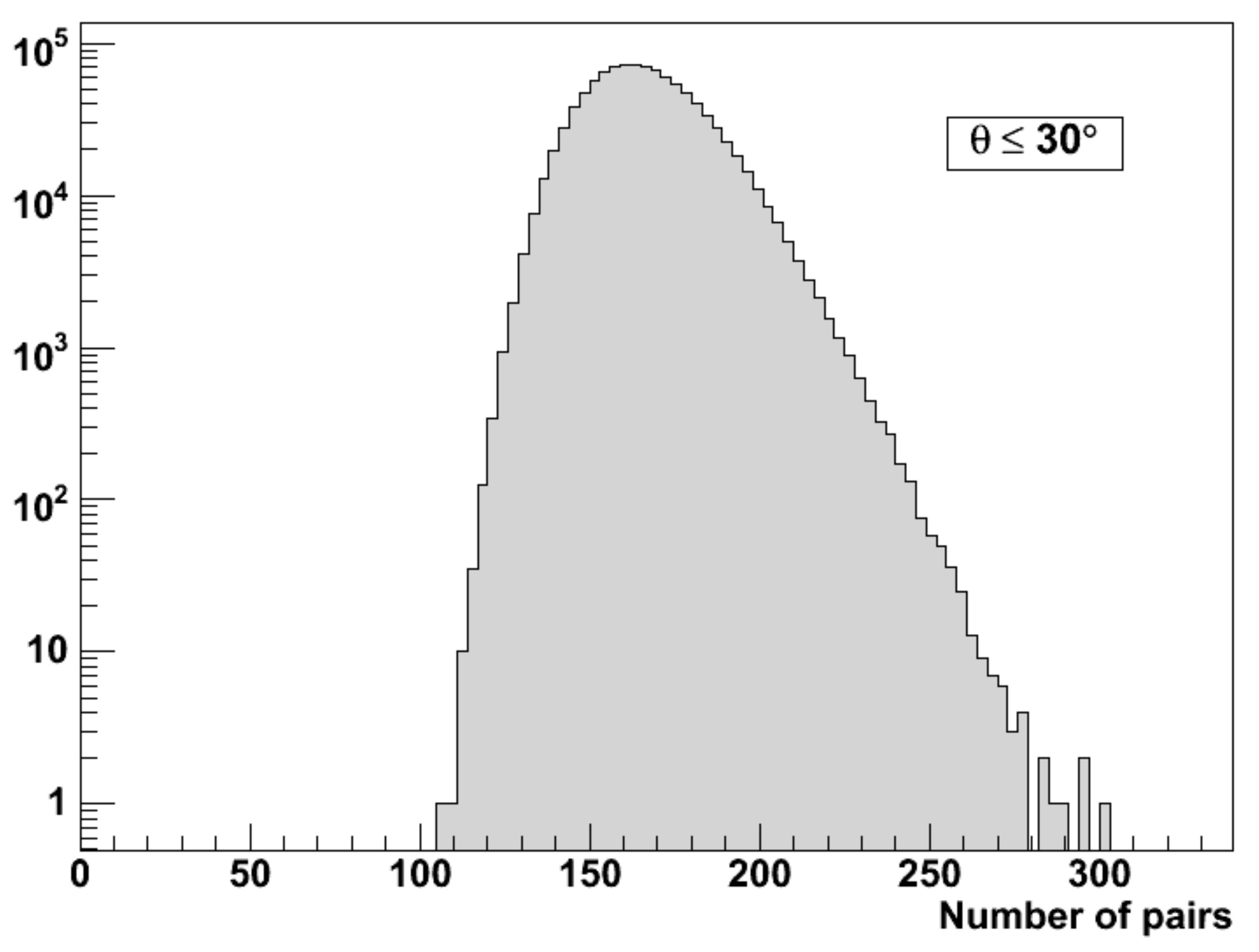}
\caption{Distributions of the number of pairs of events, $\mathcal{C}(\theta)$, within angular distance $\theta = 1^{\circ}$, $10^{\circ}$ and $30^{\circ}$, for data sets containing 60~events{ isotropically distributed,} assuming a geometrical acceptance identical to that of the Pierre Auger Observatory}
\label{fig:DPIMC}
\end{center}
\end{figure}

For each data set built in this way, we count the number of pairs of events, $\mathcal{C}(\theta)$, separated by an angular distance smaller than $\theta$, for values ranging from $1^{\circ}$ to $90^{\circ}$, by increments of $1^{\circ}$. These numbers are then compared with the numbers expected for data sets of the same size, built from tested models under similar conditions. The latter can be isotropic (Sect.~\ref{sec:sensitivity}) or correspond to generic astrophysical models of interest (Sect.~\ref{sec:doubleTest}).

\section{Sensitivity of UHECR detectors to anisotropic astrophysical models}
\label{sec:sensitivity}
The standard use of the clustering analysis consists in investigating the anisotropy of a given data set, denoted by $\mathcal{D}_{0}$, by comparing its values of $\mathcal{C}(\theta)$ with the expectations of tested data sets obtained under the assumption of an isotropic flux (weighted by the detector acceptance, see~Finley et al., 2004). To this aim, we first build the distributions of the $\mathcal{C}(\theta)$ from $10^{6}$ realizations of isotropy-based data sets. Fig.~\ref{fig:DPIMC} shows three examples of these distributions for data sets of 60 events above $E_{\min} = 60$~EeV.

\begin{figure}[!t]
\centerline{\includegraphics[width=0.9\linewidth]{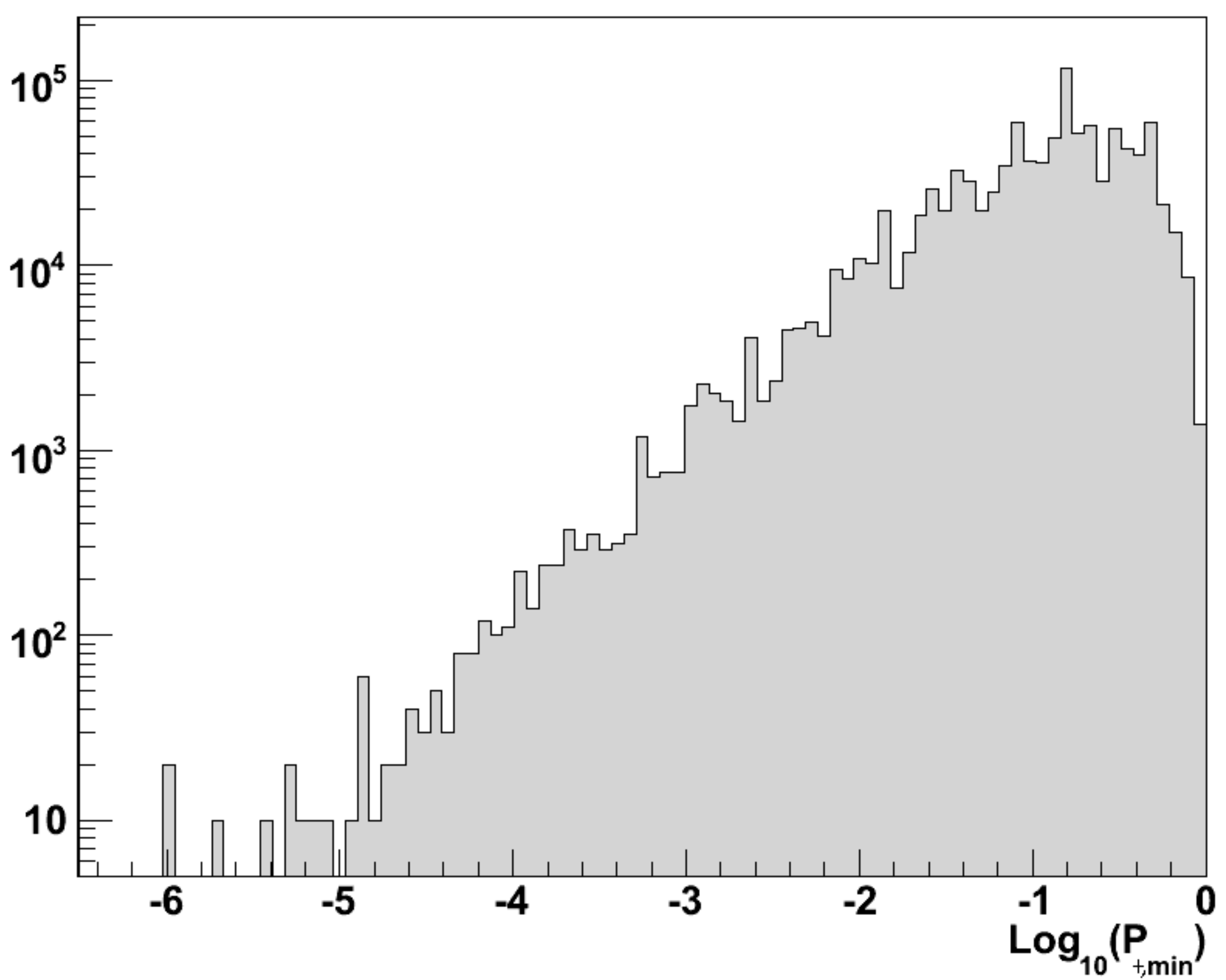}}
\caption{Distribution of the $P_{+,\min}$ values for ``isotropic'' data sets of 60 events (actually weighted by the coverage map of the Pierre Auger Observatory).}
\label{fig:Pmin}
\end{figure}

Then, for each value of $\theta$, we extract the fraction, $\mathcal{P}_{+}(\theta)$, of these ``isotropic'' data sets (the quotation marks remind one that the detector acceptance is actually taken into account) that are ``more clustered'' than the data set  $\mathcal{D}_{0}$ at angular scale $\theta$, by which we mean that they have as many or more pairs of events within $\theta$ than $\mathcal{D}_{0}$. Marginalizing over $\theta$, we then determine the minimum of this function, which occurs at some angular scale which we do not use in the current analysis. This minimum probability, $\mathcal{P}_{+,\mathrm{min}} \equiv \min\limits_{\theta}\{\mathcal{P}_{+}(\theta)\}$, is an intermediate number, which measures the level of clustering of data set $\mathcal{D}_{0}$. To turn this number into a real probability that an isotropic data set be as much (or more) clustered as $\mathcal{D}_{0}$, we finally need to complete the marginalization procedure~(see Finley et al., 2004), answering the following question: ``what is the probability that a purely random data set (in the above sense) has a value of $\mathcal{P}_{+,\mathrm{min}}$ that is lower than that found with data set $\mathcal{D}_{0}$?'' (whatever the value of $\theta$ at which this minimum occurs). For this, we apply the same procedure to another large number of isotropic data sets. This gives us the distribution of $\mathcal{P}_{+,\mathrm{min}}$ values shown in Fig.~\ref{fig:Pmin}, from which we get the fraction of these data sets which have a lower $\mathcal{P}_{+,\mathrm{min}}$. This is the so-called p-value, which we denote as $\mathcal{P}_{+}$: it gives the probability that an isotropy-based data set be ``more clustered'' than the data set under investigation, $\mathcal{D}_{0}$, at some unspecified angular scale.

\begin{figure}[!t]
\begin{center}
\hspace{-0.5cm}\includegraphics[width=0.78\linewidth,height=5.1cm]{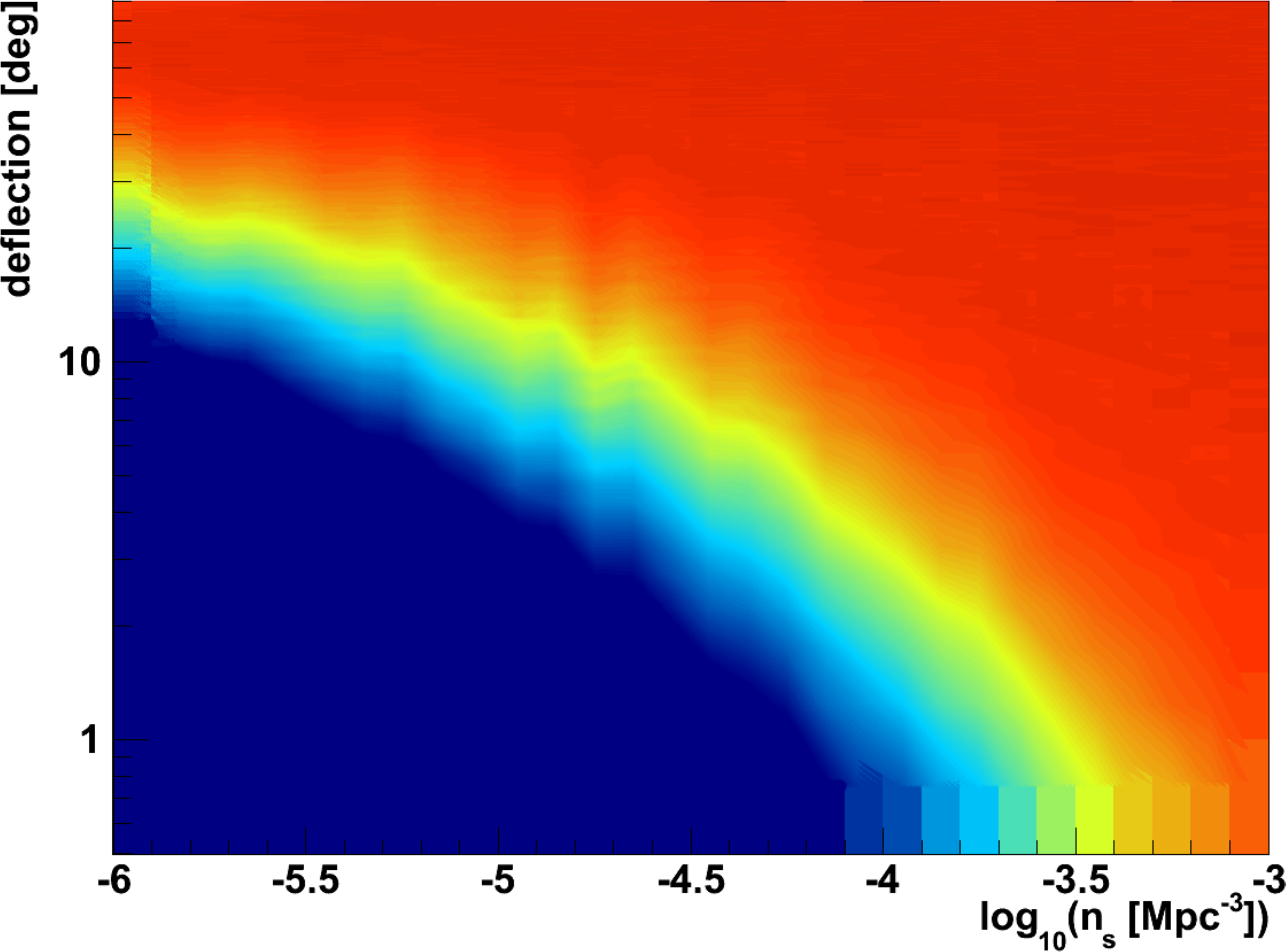}\vspace{0.05cm}\\
\hspace{-0.5cm}\includegraphics[width=0.78\linewidth,height=5.1cm]{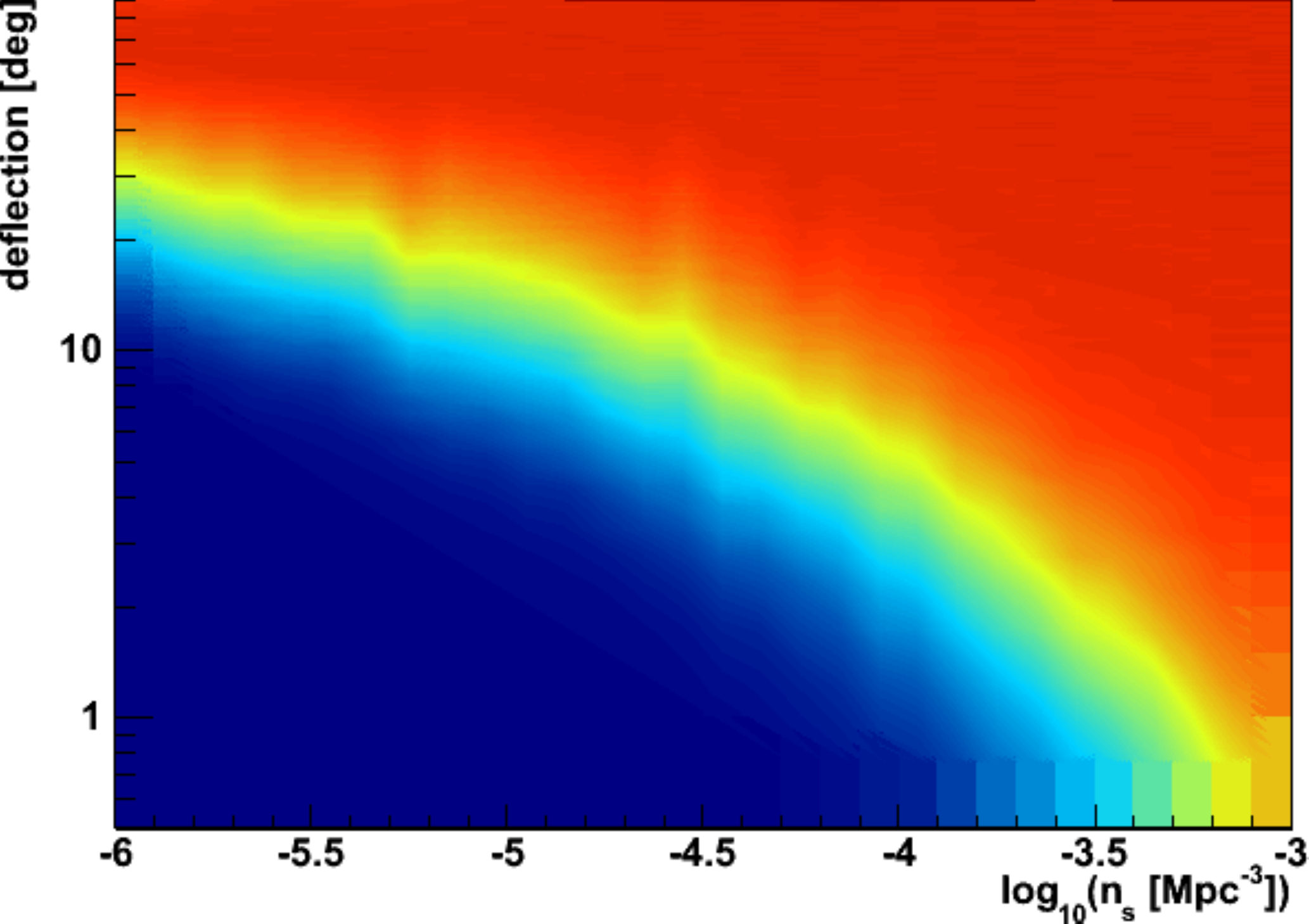}\vspace{0.05cm}\\
\hspace{-0.5cm}\includegraphics[width=0.78\linewidth,height=5.1cm]{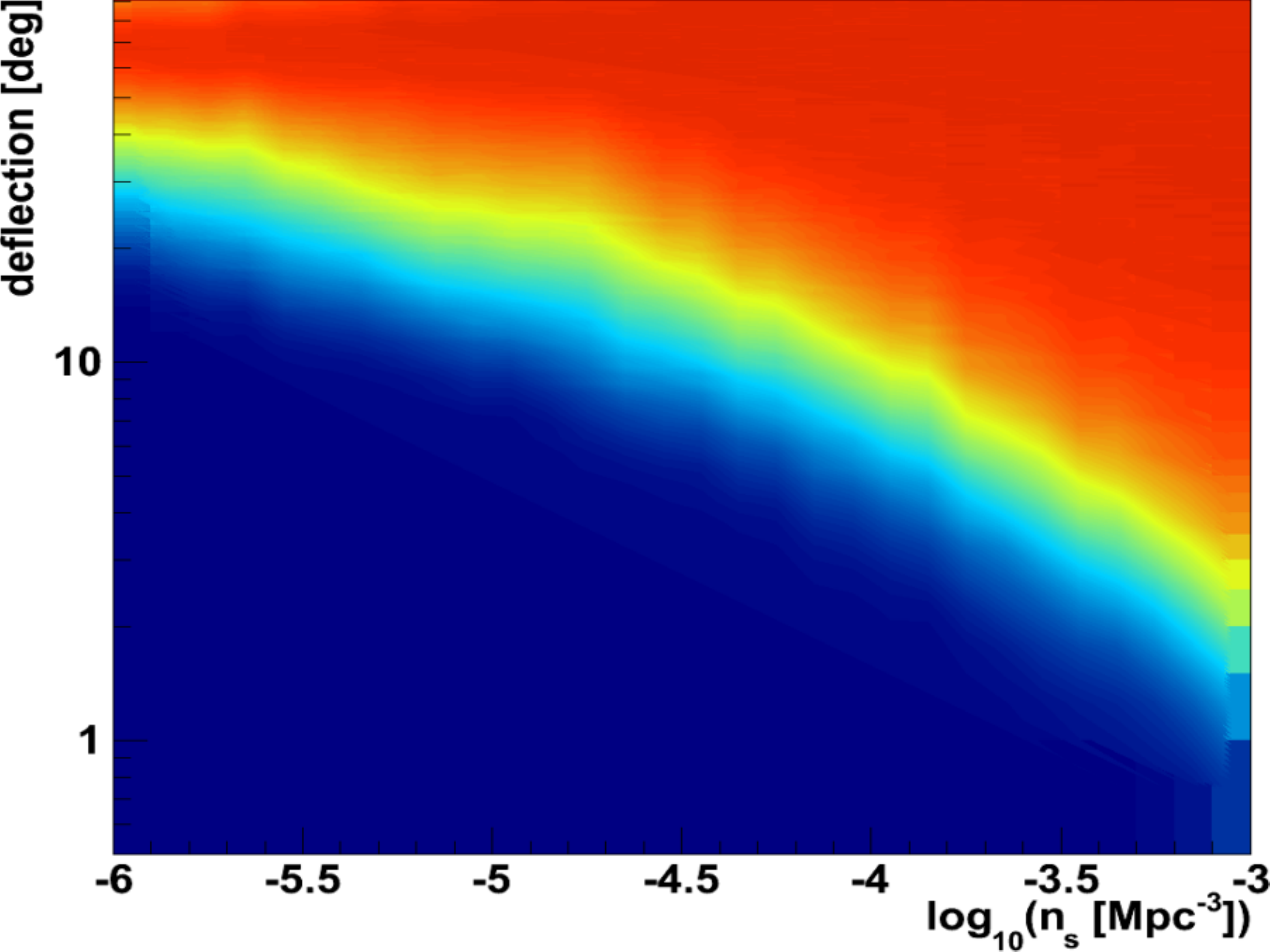}\vspace{0.05cm}\\
\hspace{0.33cm}\includegraphics[width=0.85\linewidth,height=5.2cm]{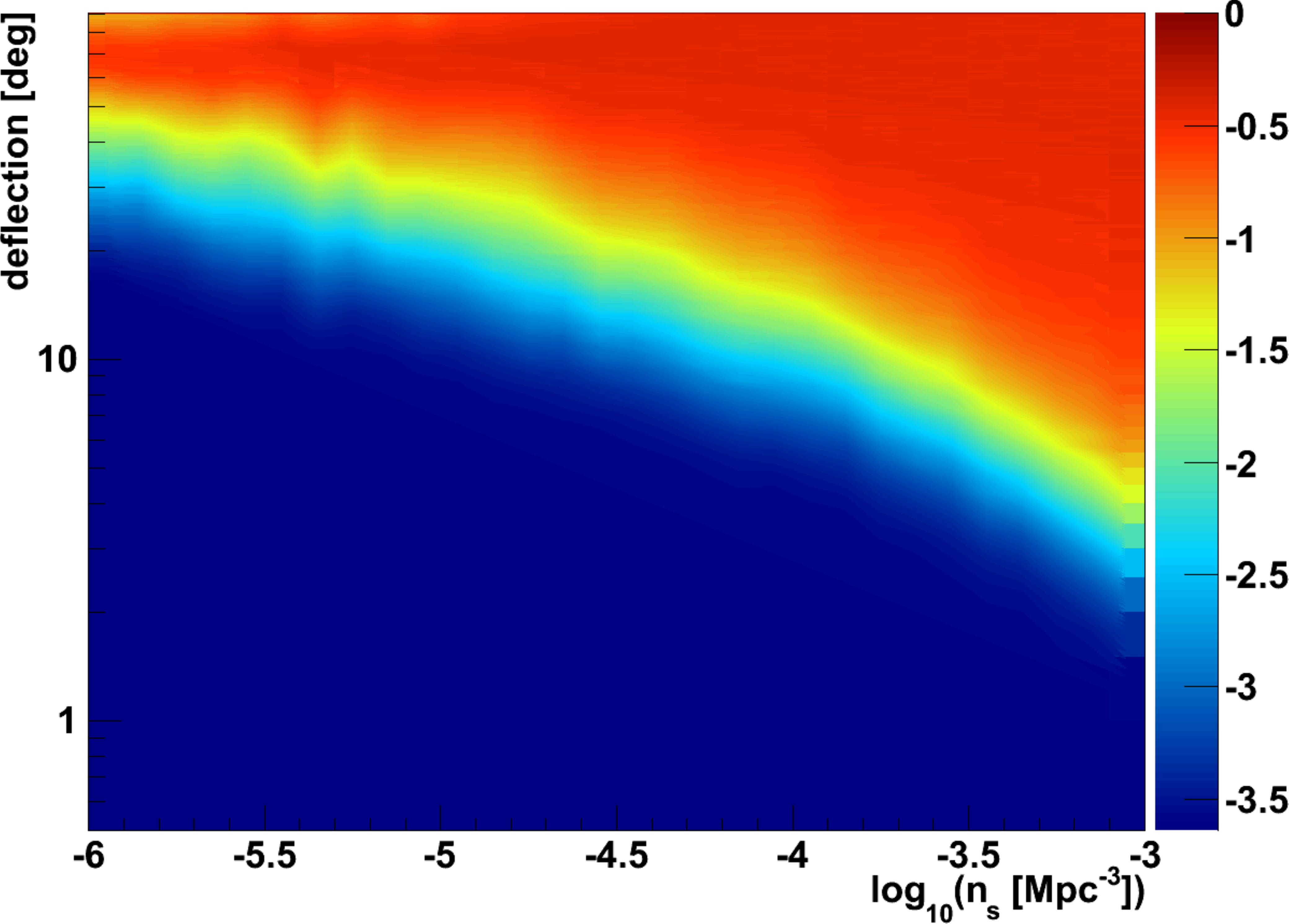}
\caption{Mean value of $\log(\mathcal{P}_{+})$ for 100 realizations of reference data sets with parameters $n_{\mathrm{s}}$ (effective density of sources, in abscissa) and $\delta$ (effective deflections, in ordinate). Four sizes of the data sets are considered: (downwards) 60, 100, 300 and 600 events. All data sets are built from an a priori uniform source distribution. {The tested model corresponds to the isotropy hypothesis.}}
\label{fig:pvalueHOM}
\end{center}
\end{figure}

\begin{figure}[!t]
\begin{center}
\hspace{-0.6cm}\includegraphics[width=0.75\linewidth,height=5.1cm]{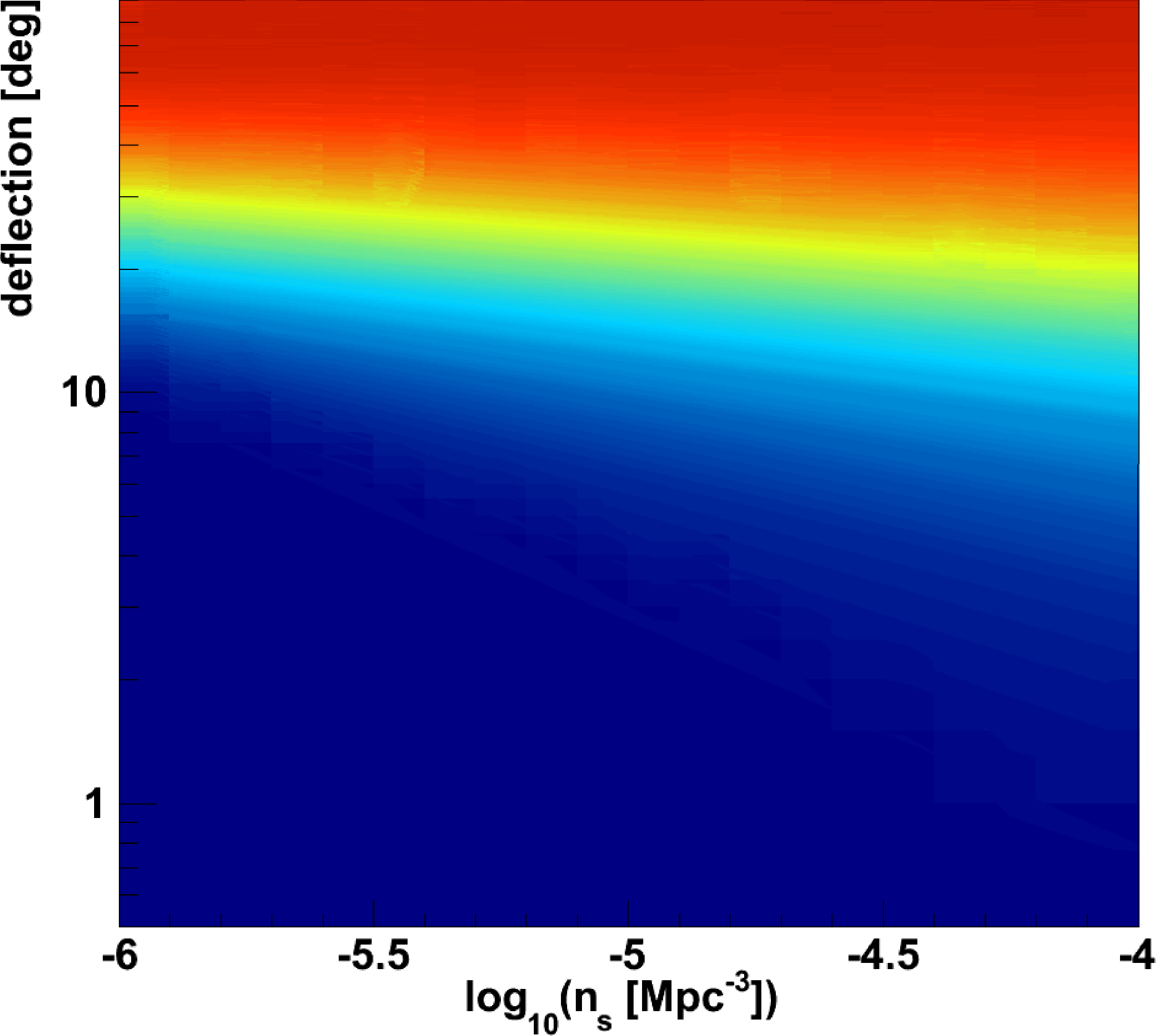}\\\hspace{-0.5cm}
\includegraphics[width=0.73\linewidth,height=5.1cm]{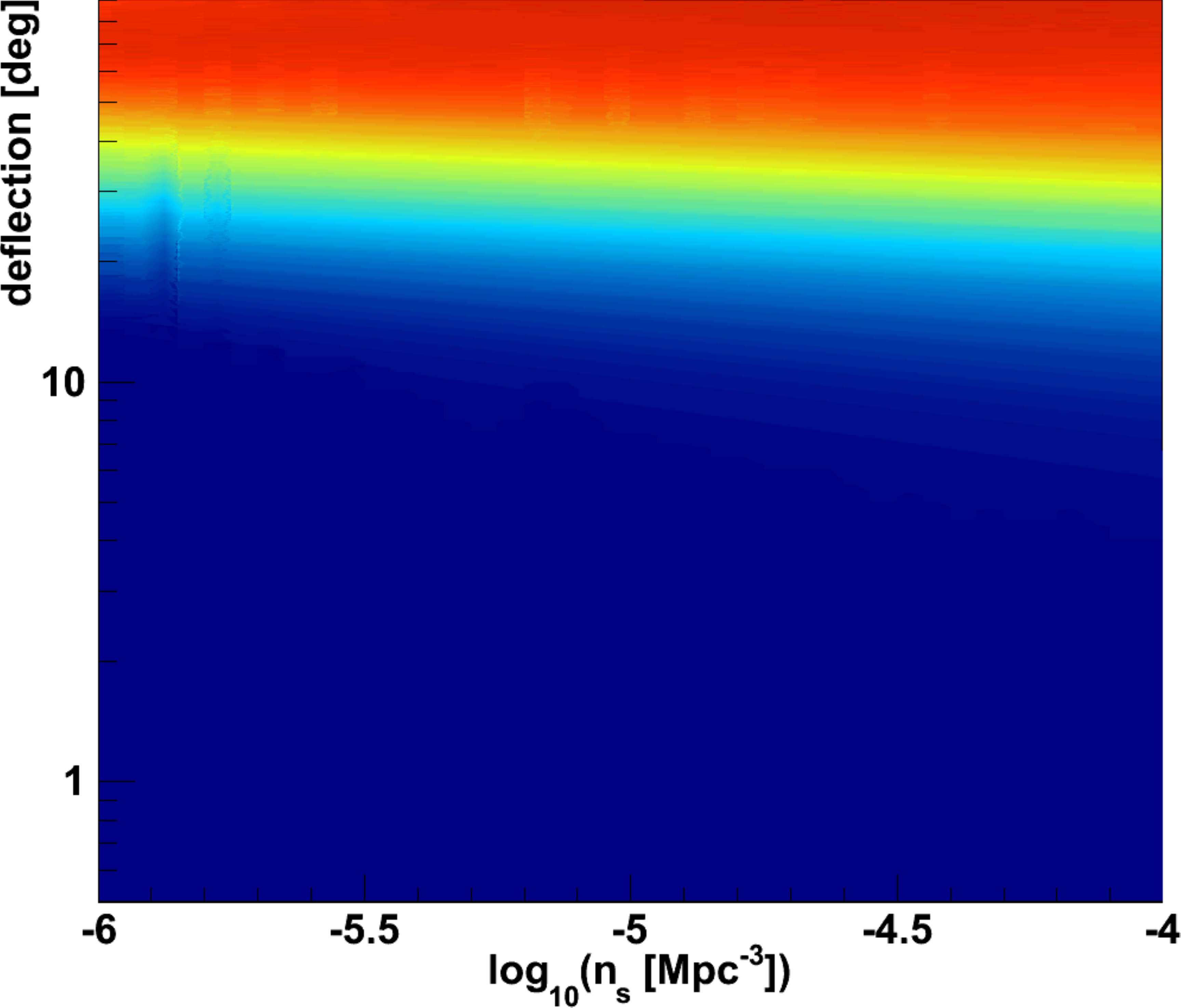}\\\vspace{0.1cm}
\hspace{-0.5cm}\includegraphics[width=0.76\linewidth,height=5.1cm]{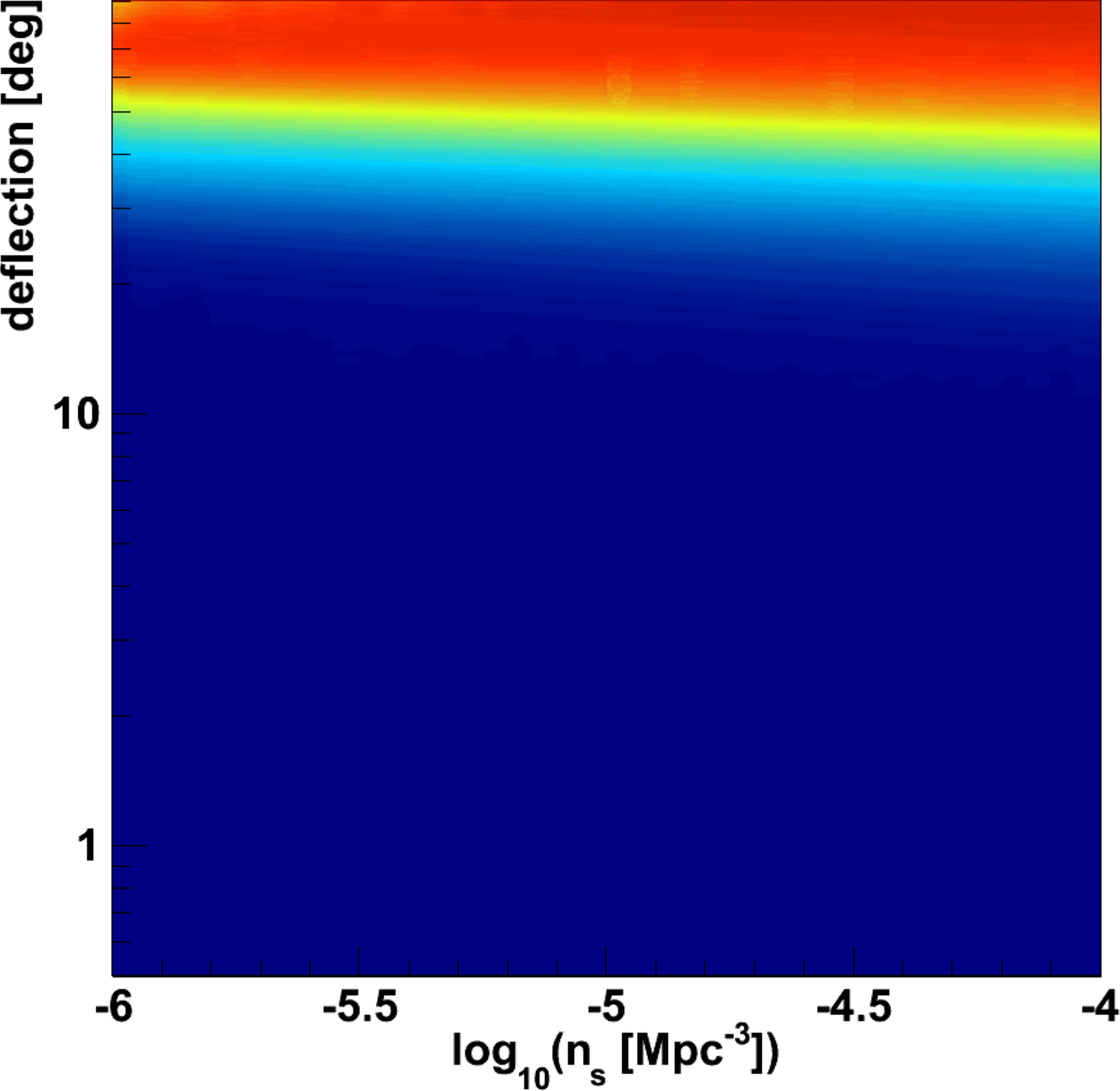}\\\hspace{0.5cm}
\includegraphics[width=0.85\linewidth,height=5.1cm]{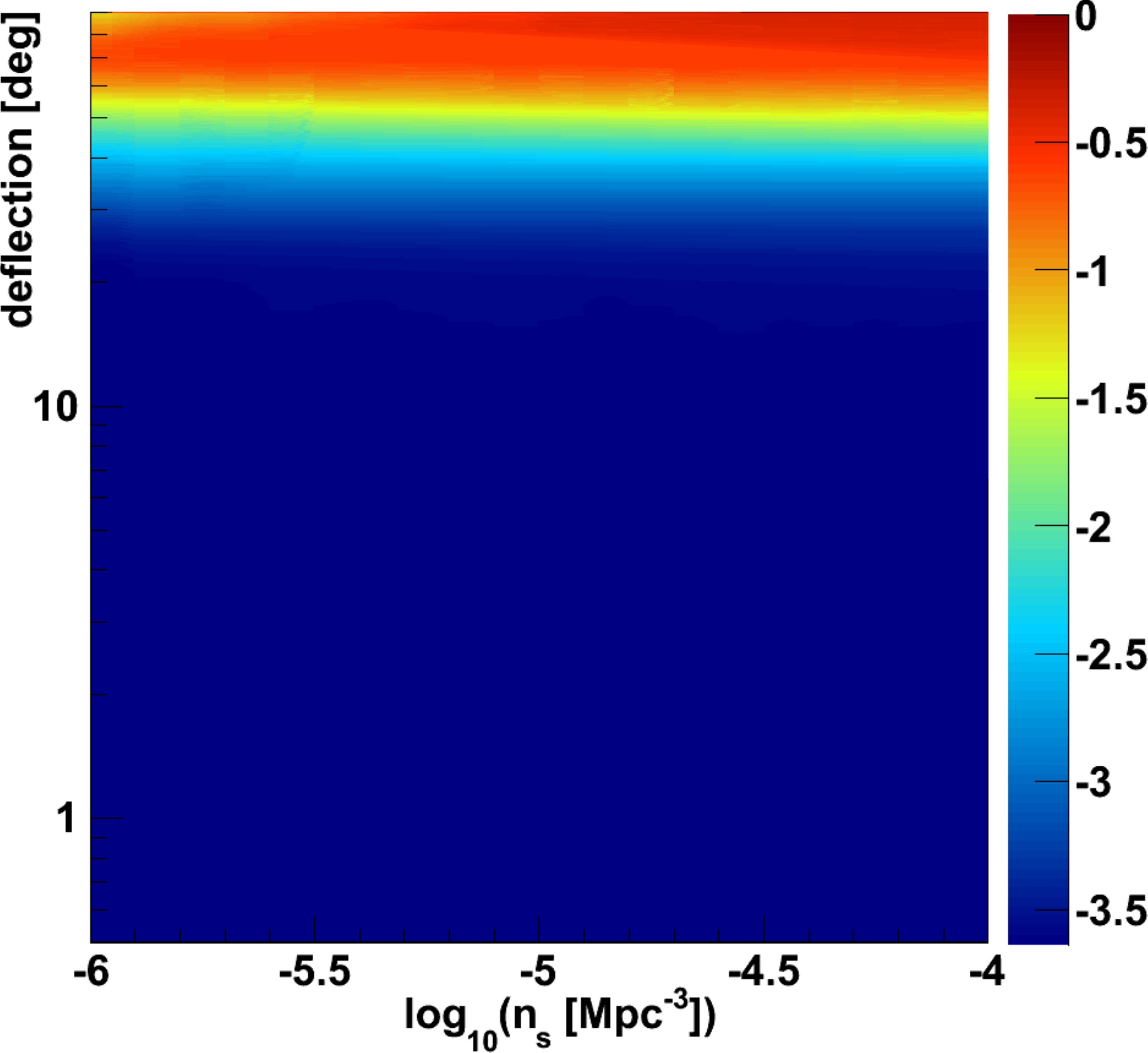}
\caption{Mean value of $\log(\mathcal{P}_{+})$ for 100 realizations of reference data sets with parameters $n_{\mathrm{s}}$ and $\delta$ (see Fig.\ref{fig:pvalueHOM} for axis definition). Four sizes of the data sets are considered: (downwards) 60, 100, 300 and 600 events. All data sets are built from an a priori anisotropic source distribution, following that of the nearby galaxies. {The tested model corresponds to the isotropy hypothesis.}}
\label{fig:pvalue2MASS}
\end{center}
\end{figure}

The usual use of this procedure is to investigate the significance of the anisotropy of a given data set. It can also be used to determine a priori the sensitivity of a given UHECR experiment to detect anisotropy, for different astrophysical models. This is illustrated in Figs.~\ref{fig:pvalueHOM} and~\ref{fig:pvalue2MASS}, where we show the mean value of $\mathcal{P}_{+}$ for 100 data sets built from astrophysical models with source density $n_{\mathrm{s}}$, shown in abscissa, and deflection scale $\delta$, shown in ordinate. In Fig.~\ref{fig:pvalueHOM}, we assume an a priori uniform source distribution in the universe, and show the sensitivity of a detector with the same coverage map as Auger, for data sets of 60, 100, 300 and 600 events above 60~EeV. As expected, the lower-left part of the plot shows a very low p-value. This is because models with few sources and small deflections will lead to strongly clustered data sets, where several events from the same source gather within a short angular scale on the sky. It is thus very unlikely that an actually isotropic sky leads to the same level of clustering. Going from left to right on the plots, the source density increases, so there are on average fewer and fewer events per source (for a given size of the data set), and the clustering signal becomes smaller and smaller, even for a low deflection scale. In the limit where there is never more than one event per source, the observed clustering of the events reflects the clustering of the sources themselves. Going from bottom to top, the angular deflection scale increases, so the clustering of events from the same source is blurred over larger scales, and if the sources are numerous enough (large source density), their individual ``image'' on the sky will overlap, making the data set more and more similar to a data set drawn from an isotropic flux. This explains the large p-value found in the upper-right part of the plots. Finally, for larger data sets, there will be on average more events per source (whatever the source density), so the intrinsic clustering of the UHECR events will be larger, compared to the isotropic expectations. This is why the low p-value regions are basically moved towards the upper-right corner as the data set becomes larger, as seen on the figures.

These plots can be regarded as indicating the range of astrophysical models for which the detector is sensitive enough to detect a significant anisotropy in data sets of a given size, as represented by the blue-green part of the parameter space. For instance, if a detector can collect 100 events above 60~EeV, it can be expected to assess the anisotropy of the UHECR sky at the 99.9\% confidence level (C.L.) for models with a source density lower than $2\,10^{-4}\,\mathrm{Mpc}^{-3}$ and a deflection scale of the order of 1$^{\circ}$ or models with a source density around $10^{-6}\,\mathrm{Mpc}^{-3}$ and a deflection scale smaller than 15$^{\circ}$. The sensitivity is clearly improved for detectors able to collect 600 events above the same energy (see below for further comments).

{In Fig.~\ref{fig:pvalue2MASS}, we show the same results for the hypothesis of an a priori source distribution similar to the distribution of the galaxies in the local universe, taken as a subsample of the 2MRS catalog (Huchra et al., 2011). The maximum distance of the sources is $\sim$200 Mpc, taken as the GZK horizon for 60-EeV-protons (see above). Moreover,  the subsample completeness is ensured by cutting out sources with a magnitude in the K-band lower than 11.25\footnote{{These cuts lead to an intrinsic density of the subsample of $\sim10^{-3.4}$ Mpc$^{-3}$. Due to limited CPU time, we limited the scan of the density/deflection space to deflections of up to 10$^{-4}$ Mpc$^{-3}$.}}.} As is shown in Fig.~\ref{fig:pvalue2MASS}, the same trends as in the hypothesis of a distribution are observed. The main difference is that the models assuming small UHECR deflections appear incompatible with isotropic expectations even in the case of large source densities (lower-right part of the figures). The reason for that is that, even when there is only one event per source (small data sets), the clustering of the sources themselves is apparent in the resulting data sets, producing a significant clustering signal that isotropic data sets can only exceptionally exhibit by chance.

This result is interesting because it shows that detectors with moderate acceptance can detect a significant anisotropy signal in a large number of astrophysical cases, or, failing to detect this, constrain a large class of astrophysical models, typically rejecting the blue area. Indeed, the data sets that a detector can collect if the actual source density and cosmic ray deflections correspond to a point in this region will generally be more clustered than 99.9\% of the isotropy-based data sets.

For example, let us consider a data set of $N_{\mathrm{evt}}$ events above 60 EeV that would show an auto-correlation signal at the level of 1\%. From the point of view of this clustering analysis, such a data set would thus be similar to the data sets collected from the astrophysical models corresponding to the green area in the figures. Note that this does not mean that the actual universe \emph{must be} one of these models. One can only say that the data set collected from the actual universe is similar to the typical data sets expected for these models, which could occasionally be obtained from other types of models. So to improve the analysis, it would be interesting to investigate the probability that a given type of astrophysical models lead to a data set similar to the data set under investigation (from the point of view of its auto-correlation). This is addressed in the next section.

\begin{figure}[!t]
\begin{center}
\includegraphics[width=0.85\linewidth]{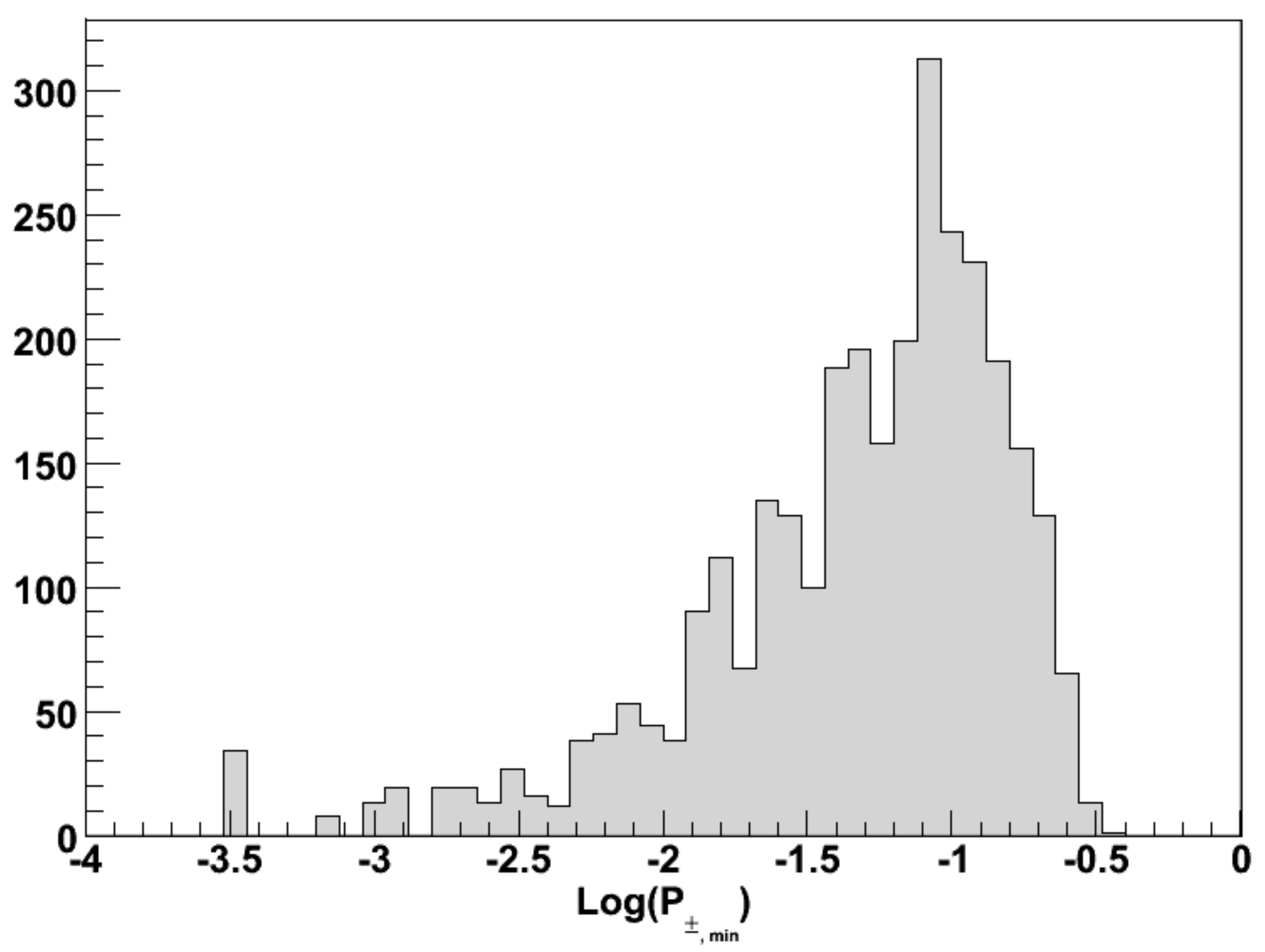}
\caption{Distribution of the values of $\mathcal{P}_{\pm,\min}$ for 3000 data sets of 60 UHECR events above 60~EeV, obtained with the following test model: $n_{\mathrm{s}} = 10^{-4}\,\mathrm{Mpc}^{-3}$, $\delta = 15^{\circ}$, and a uniform a priori source distribution.}
\label{fig:DistribPminMoreOrLess}
\end{center}
\end{figure}

\section{Deriving constraints on astrophysical models
\label{sec:doubleTest}}

\subsection{Comparing data with classes of models}

In order to derive astrophysically meaningful constraints from the clustering analysis of a given reference data set, $\mathcal{D}_{0}$, we now wish to compare its clustering signal with that of tested models which are no longer isotropic, but span the whole parameter space ($n_{\mathrm{s}}$, $\delta$ and the type of source distribution). We shall thus propose a general test of a given astrophysical hypothesis, to be tested on the data set $\mathcal{D}_{0}$.

The data set $\mathcal{D}_{0}$ has two opposite ways to differ from the expectations of a given astrophysical model, $\mathcal{M}_{\mathrm{test}}$: it can ``more clustered'' or ``less clustered'' than the typical data sets produced by this model. To quantify this potential discrepancy, we compute for each value of $\theta$ the distributions of $\mathcal{C}(\theta)$, defined in Sect.~\ref{sec:ctheta}, for $10^{4}$ independent realizations of data sets of the same size as $\mathcal{D}_{0}$, built within model $\mathcal{M}_{\mathrm{test}}$. These distributions are similar to those shown in Fig.~\ref{fig:DPIMC}. We then determine, for each $\theta$, the fraction $\mathcal{P}_{+}(\theta)$, resp. $\mathcal{P}_{-}(\theta)$, of tested data sets that have higher, resp. lower, values of $\mathcal{C}(\theta)$ than data set $\mathcal{D}_{0}$. From these, we obtain the angular scale where $\mathcal{D}_{0}$ appears most unlike the typical tested models, whether because it is more clustered or less clustered. We note the corresponding fraction $\mathcal{P}_{\pm,\min}$:
\begin{equation}
\mathcal{P}_{\pm,\min} = \min\limits_{\theta}\{\mathcal{P}_{+}(\theta),\mathcal{P}_{-}(\theta)\}.
\label{eq:ppm}
\end{equation}

Finally, we proceed as in the previous section and determine which fraction of another large sample of data sets built within the same model $\mathcal{M}_{\mathrm{test}}$ have a smaller value of $\mathcal{P}_{\pm,\min}$ than the one obtained with $\mathcal{D}_{0}$. This number is noted $\mathcal{P}_{\pm}$, and for simplicity, we call it the \emph{clustering similarity} of data set $\mathcal{D}_{0}$ with respect to model $\mathcal{M}_{\mathrm{test}}$ (see below for its precise meaning).

\begin{figure}[!h]
\begin{center}
\includegraphics[width=0.9\linewidth,height=6cm]{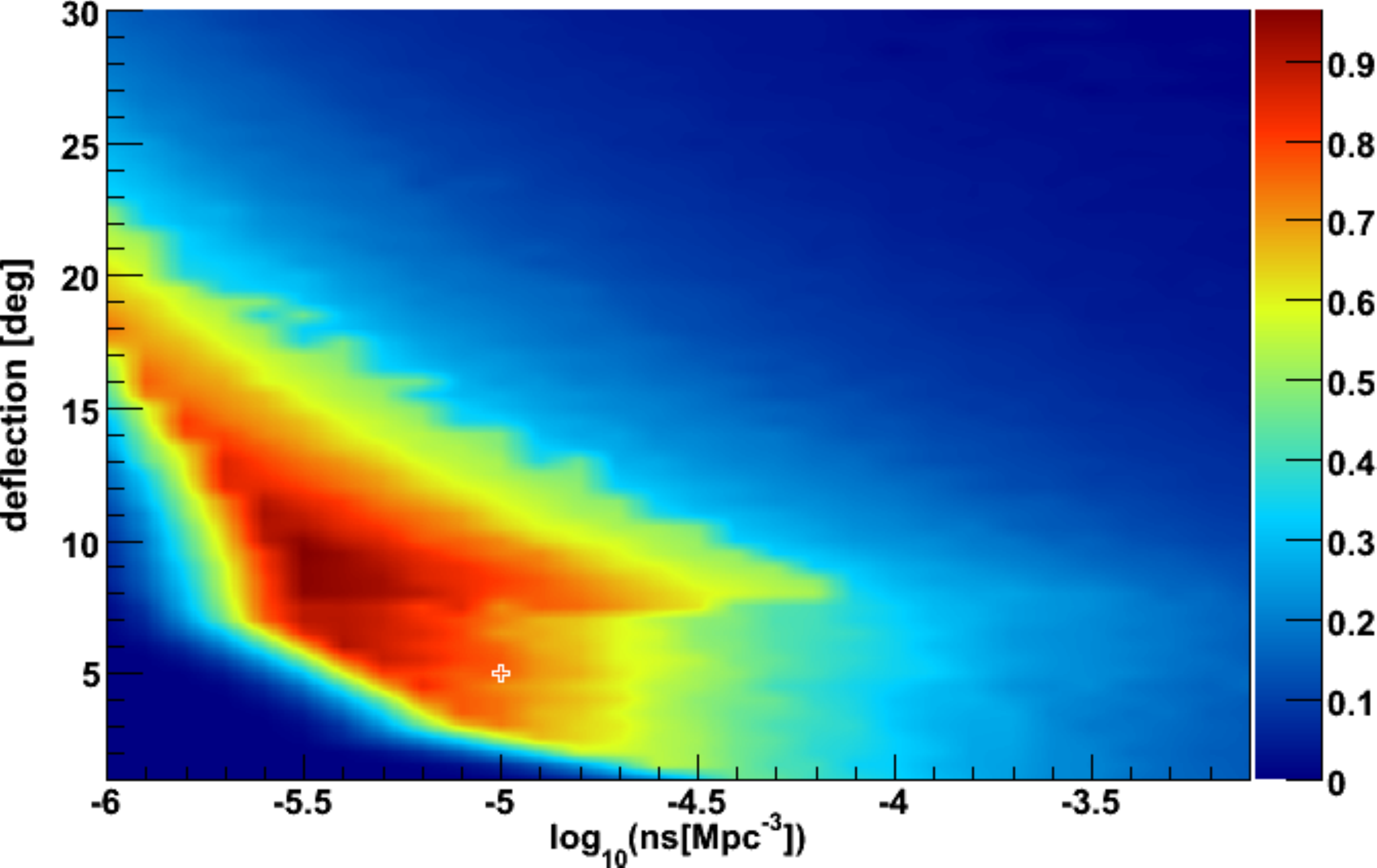}\\
\hspace{-0.7cm}\includegraphics[width=0.812\linewidth,height=6cm]{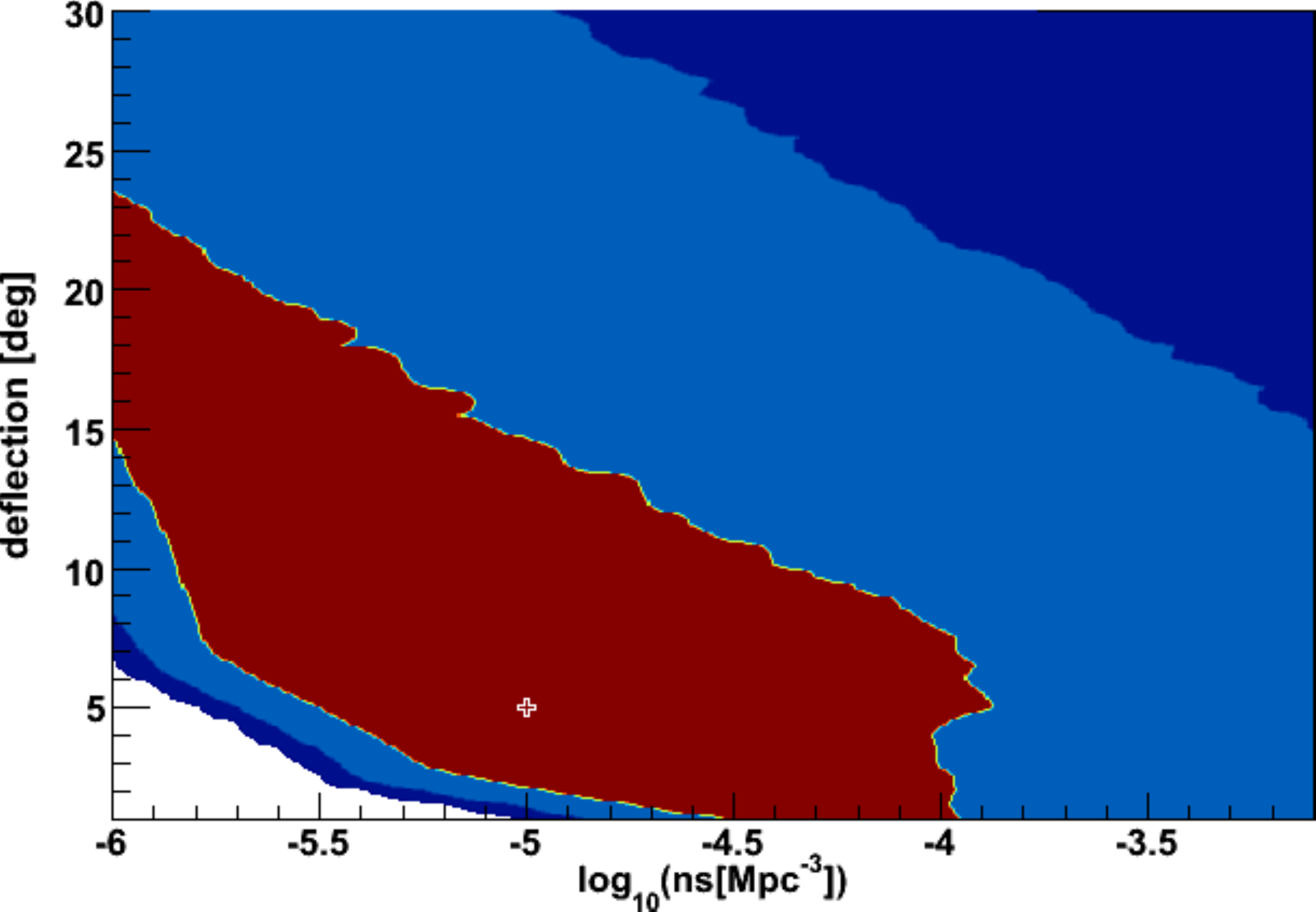}\hfill
\caption{Top: values of the ``clustering similarity'' $\mathcal{P}_{\pm}$ for all tested pairs of parameters $(n_{\mathrm{s}},\delta)$, obtained from the analysis of a typical data set built from a realization of the reference model $\mathcal{D}_{0}$ = $(n_{\mathrm{s}} = 10^{-5}\,\mathrm{Mpc}^{-3},\delta = 5^{\circ})$, indicated by the white cross. {For all models considered here, a uniform distribution of the sources was assumed.} Bottom: confidence contours of 68\% (red), 95\% (light blue) and 99\% (dark blue) extracted from the same data set $\mathcal{D}_{0}$. These plots can be interpreted as giving the regions of the parameter space that are most compatible with the reference data set (see text for details).}
\label{fig:likely_50_5_5_emin_60EeV}
\end{center}
\end{figure}

As an example, Fig.~\ref{fig:DistribPminMoreOrLess} shows the distribution of the values of $\mathcal{P}_{\pm,\min}$ for 3000 data sets of 60 UHECR events above 60~EeV, obtained with the following test model: $n_{\mathrm{s}} = 10^{-4}\,\mathrm{Mpc}^{-3}$, $\delta = 15^{\circ}$, uniform a priori source distribution. Then $\mathcal{P}_{\pm}$ is simply the fraction of the data sets that appear on the left side of the value of $\mathcal{P}_{\pm,\min}$ obtained for the data set $\mathcal{D}_{0}$ in this histogram:
\begin{equation}
\mathcal{P}_{\pm} = \int_{-\infty}^{\mathcal{P}_{\pm,\min}(\mathcal{D}_{0})}f(\mathcal{P}_{\pm,\min})\mathrm{d}\mathcal{P}_{\pm,\min},
\label{eq:PPlusMinusMin}
\end{equation}
where $f(\mathcal{P}_{\pm,\min})$ is the function that represents the normalized histogram.

So $\mathcal{P}_{\pm}$ is the probability that the test model, $\mathcal{M}_{\mathrm{test}}$, leads to a data set as dissimilar or more dissimilar to the average clustering features corresponding to this model than the data set under investigation, $\mathcal{D}_{0}$. 

To illustrate how this method can be used on a real data set, we choose a particular astrophysical model in the parameter space and generate from this model one data set, $\mathcal{D}_{0}$, to be analyzed through its clustering properties. To generate Fig.~\ref{fig:likely_50_5_5_emin_60EeV}, we started with a realization of an a priori uniform source model, with a source density $n_{\mathrm{s}} = 10^{-5}\,\mathrm{Mpc}^{-3}$ and a deflection angular scale $\delta = 5^{\circ}$, as indicated by the white cross on the plot. We then propagated cosmic rays from this realization of the sources to build a data set of $N_{\mathrm{evt}} = 50$~events above $E_{\min} = 60$~EeV, which is our reference data set $\mathcal{D}_{0}$.

The colors in Fig.~\ref{fig:likely_50_5_5_emin_60EeV} indicate the value of $\mathcal{P}_{\pm}$ for $\mathcal{D}_{0}$. The region of the parameter space appearing in blue corresponds to models which have a low probability to yield data sets with clustering properties similar to that of $\mathcal{D}_{0}$ (namely clustering properties at least as different as those of $\mathcal{D}_{0}$ from the typical clustering properties for these models). By contrast, the regions in red correspond to models which yield data sets with clustering properties similar to $\mathcal{D}_{0}$. On the lower-left of the plot, the data sets are typically \emph{much more} clustered (i.e. show more pairs within small angular distances) than $\mathcal{D}_{0}$. On the upper-right, on the contrary, the typical data sets are \emph{much less} clustered than $\mathcal{D}_{0}$, because there are more sources contributing, each of which appears through an image broadened by larger deflections. This illustrates the power of an analysis that combines both the excesses and deficits of clustering of the data set under investigation, with respect to data sets produced by different astrophysical models. It can do more than simply assessing the anisotropy of the data set: it can point towards candidate astrophysical models.

\subsection{Comments on the statistical meaning of the ``clustering similarity''}

To make this statement more precise, from the statistical point of view, it is interesting to indicate how confidence intervals can be derived for the model parameters. The values of the clustering similarity $\mathcal{P}_{\pm}$ given in Fig.~\ref{fig:likely_50_5_5_emin_60EeV} for each $n_{\mathrm{s}}$ and $\delta$ are essentially p-values in a goodness-of-fit approach. This, of course, does not give the probability of the different hypotheses. The plot can however be read in a Bayesian perspective as giving the posterior probability density function (p.d.f.), after normalization over the whole parameter space, for an implicit prior p.d.f. which is flat in deflection angular scales and flat in the logarithm of the source density, in the explored range.

In a frequentist approach, we can also adapt the Neyman construction for the confidence intervals to the specific case of our clustering analysis. The situation is slightly more subtle than in usual cases, because here the ``measured observable'', which is $\mathcal{P}_{\pm,\min}$, is not derived from the data alone, but depends on the tested model as well. The actual values measured by the experiment are the number of pairs within each angular scale $\theta$, that is $\mathcal{C}(\theta)$. From this, we obtained a ``relatively measured observable'' $\mathcal{P}_{\pm,\min}(n_{\mathrm{s}},\delta)$ for each tested hypothesis, by which we mean that, once a set of parameters $(n_{\mathrm{s}},\delta)$ is chosen for the reference data sets, $\mathcal{P}_{\pm,\min}$ can be univocally computed from the data, and is thus an observable in its own right. In the above sense, however, it is measured \emph{relatively to the model} $(n_{\mathrm{s}},\delta)$.

The Neyman procedure can nevertheless be applied. For a pre-specified \emph{confidence level}, $1 - \alpha$, we need to determine limiting values $\mathcal{P}_{\pm,\min,1}$ and $\mathcal{P}_{\pm,\min,2}$ such that
\begin{equation}
\int_{\mathcal{P}_{\pm,\min,1}}^{\mathcal{P}_{\pm,\min,2}}f(\mathcal{P}_{\pm,\min};n_{\mathrm{s}},\delta)\mathrm{d}\mathcal{P}_{\pm,\min} = 1 - \alpha,
\label{eq:Neyman1}
\end{equation}
where $f(\mathcal{P}_{\pm,\min};n_{\mathrm{s}},\delta)$ is the density of probability that one particular data set computed with the assumption that the source density is $n_{\mathrm{s}}$ and the deflection scale is $\delta$, be characterized by the value $\mathcal{P}_{\pm,\min}$ with respect to this choice of parameters. In other words, $f(\mathcal{P}_{\pm,\min};n_{\mathrm{s}},\delta)$ is simply the (normalized) histogram shown in Fig.~\ref{fig:DistribPminMoreOrLess} (of which there is one for each point of the parameter space).

Now we can choose to set $\mathcal{P}_{\pm,\min,2} = 1$, in which case $\mathcal{P}_{\pm,\min,1}$ satisfies:
\begin{equation}
\int_{-\infty}^{\mathcal{P}_{\pm,\min,1}}f(\mathcal{P}_{\pm,\min};n_{\mathrm{s}},\delta)\mathrm{d}\mathcal{P}_{\pm,\min} = \alpha.
\label{eq:Neyman2}
\end{equation}
So, comparing with Eq.~(\ref{eq:PPlusMinusMin}), one sees that $\mathcal{P}_{\pm,\min,1}$ is identical to $\mathcal{P}_{\pm,\min}(\mathcal{D}_{0})$ for all the points in the parameter space that give a p-value $\mathcal{P}_{\pm}$ precisely equal to $\alpha$ in the plot of Fig.~\ref{fig:likely_50_5_5_emin_60EeV}. In other words, the contour defined by the level curve (i.e. identical color contour) with value $\alpha$ in the plot also defines a contour of values of $\mathcal{P}_{\pm,\min,1}$ for the particular choice of confidence level $1 - \alpha$.

Thus, given the Neyman analysis of the confidence intervals, this also means that, by construction, the level contours of the $\mathcal{P}_{\pm}$ plots, such as Fig.~\ref{fig:likely_50_5_5_emin_60EeV}, are exactly the Neyman frequentist confidence contours, corresponding to frequencies equal to $1-\mathcal{P}_{\pm}$. In other words, starting with a data set $\mathcal{D}_{0}$ built from a given astrophysical model $(n_{\mathrm{s}},\delta)$, we can apply the above procedure to draw the clustering similarity $\mathcal{P}_{\pm}$ based on this data set, and these contours will encircle the actual model from which the data set was built in exactly a fraction $1-\mathcal{P}_{\pm}$ of the cases.

As an illustration, we show in Fig.~\ref{fig:likely_50_5_5_emin_60EeV} the confidence contours extracted from the same analyzed data set $\mathcal{D}_{0}$ as the one used to draw the Bayesian p.d.f. If one repeats the construction of these contours over many realizations of the same model ($n_{\mathrm{s}} = 10^{-5}\,\mathrm{Mpc}^{-3}$, $\delta=5^{\circ}$, as indicated by the white cross), then by construction the red-orange region will enclose the white cross in 68\% of the cases. Note that, in Fig.~\ref{fig:likely_50_5_5_emin_60EeV}, we chose a realization where this property was actually fulfilled. 

\begin{figure}[!t]
\begin{center}
\includegraphics[width=0.85\linewidth,height=6.5cm]{Contours_50_5_5_emin_60EeV.pdf}\\
\hspace{-0.13cm}\includegraphics[width=0.9\linewidth,height=6.9cm]{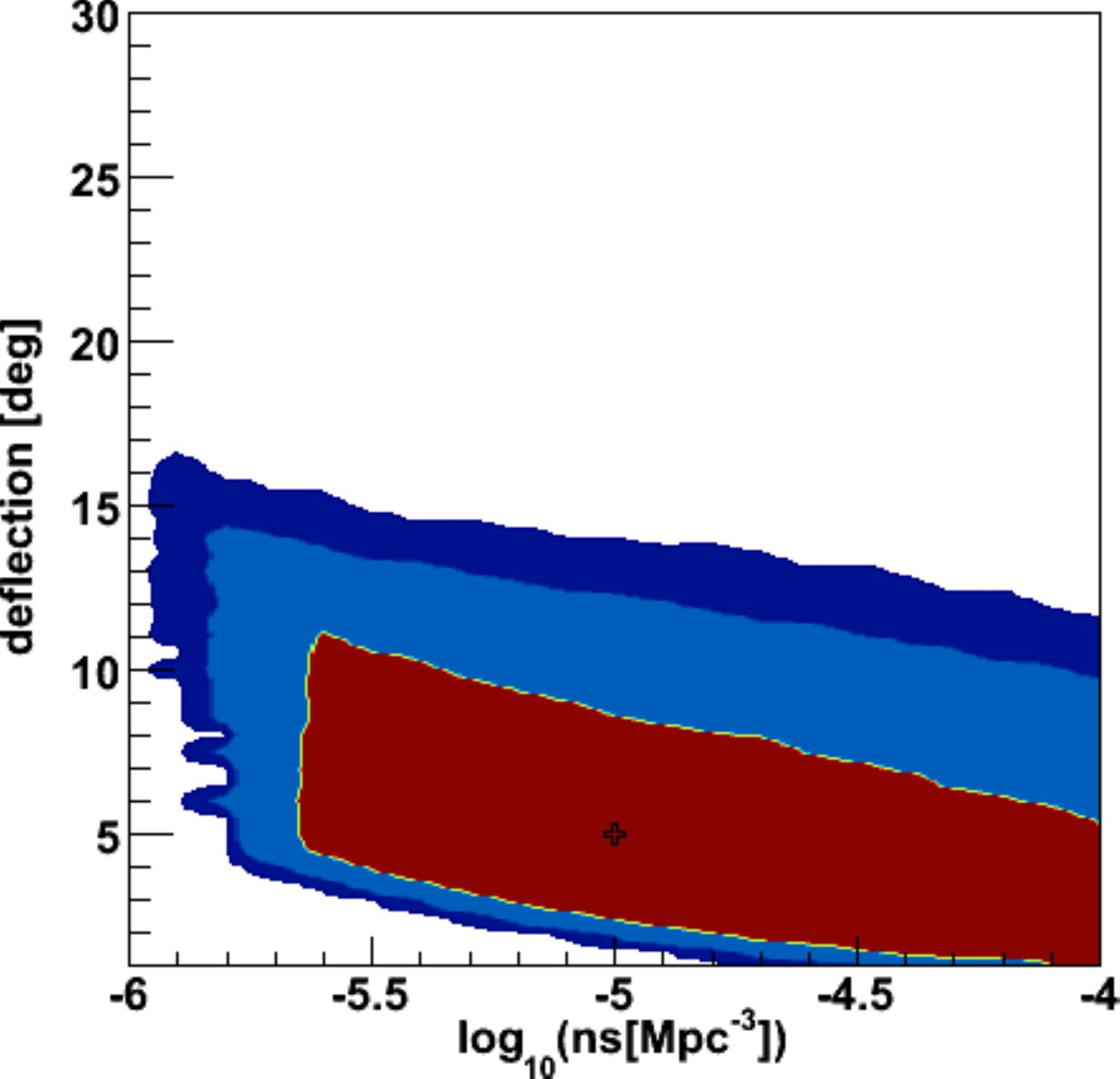}
\caption{Influence of the underlying source distribution on the astrophysical constraints derived from the clustering analysis. The confidence contours (see~Fig.~\ref{fig:likely_50_5_5_emin_60EeV} for color meaning) are extracted from the analysis of a typical data set of 50 events above an energy threshold of 60~EeV, built from a typical realization of the reference model $(n_{\mathrm{s}} = 10^{-5}\,\mathrm{Mpc}^{-3},\delta = 5^{\circ})$, indicated by a cross, assuming either an a priori uniform source distribution (on the left) or an anisotropic distribution (on the right) similar to that of the galaxies (see text).}
\label{fig:conts_distrib}
\end{center}
\end{figure}

\subsection{Influence of the parameters}

\subsubsection{Spatial distribution of UHECR sources}

The influence of the source distribution on the ability of a UHECR experiment to detect an anisotropy through our clustering analysis has already been shown in Figs.~\ref{fig:pvalueHOM} and~\ref{fig:pvalue2MASS}. On Fig.~\ref{fig:conts_distrib}, we further show that a prior assumption on the spatial distribution of the UHECR sources modifies the astrophysical constraints that can be drawn from an anisotropic data set. In this example, we choose the same density/deflection parameters as above (namely $n_{\mathrm{s}} = 10^{-5}\,\mathrm{Mpc}^{-3}$ and $\delta=5^{\circ}$), both with an a priori uniform source distribution (on the left) and with a source distribution following the 2MRS galaxy distribution, which is intrinsically anisotropic (on the right). Clearly, a prior expectation regarding the overall source distribution (here similar to the galaxies in the nearby universe) helps reducing the range of parameters compatible with the clustering properties of the data set under investigation. 

\subsubsection{Size of the data set}

\begin{figure*}[!t]
\begin{center}
\includegraphics[width=0.42\linewidth,height=6.6cm]{Likely_50_5_5_emin_60EeV.pdf}\;\;\;
\includegraphics[width=0.42\linewidth,height=6.6cm]{Contours_50_5_5_emin_60EeV.pdf}\\
\includegraphics[width=0.42\linewidth,height=6.6cm]{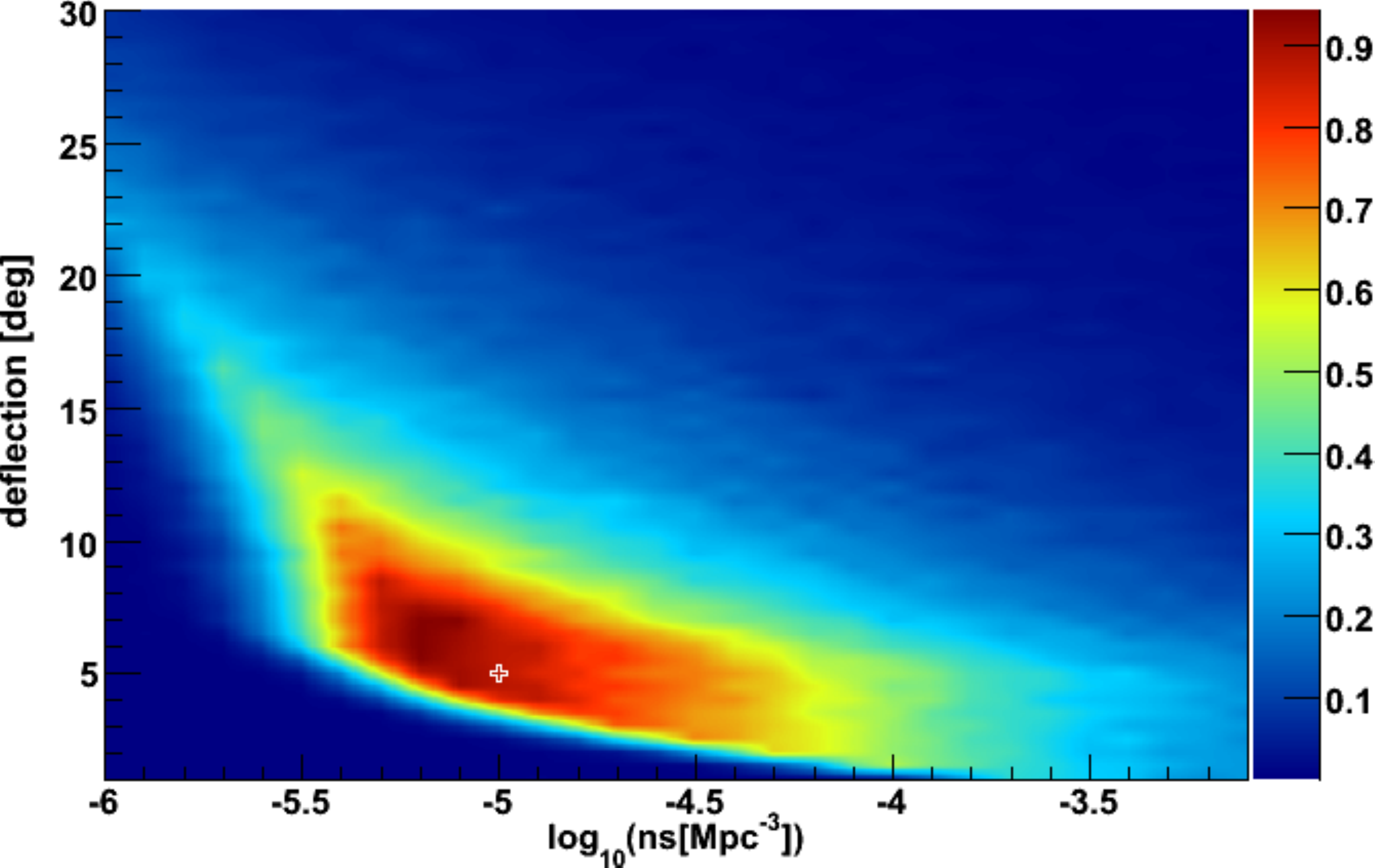}\;\;\;
\includegraphics[width=0.42\linewidth,height=6.6cm]{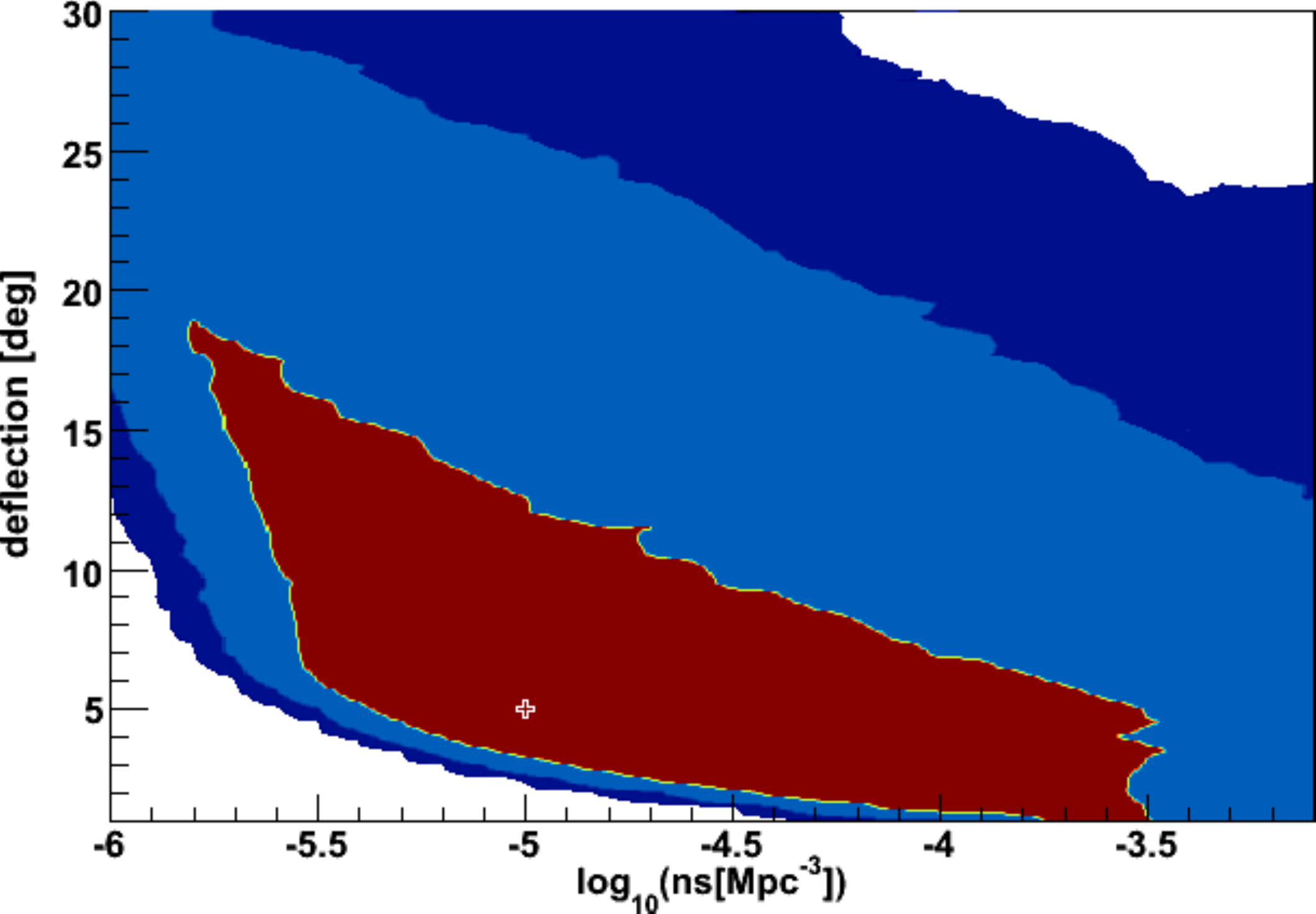}\\
\includegraphics[width=0.42\linewidth,height=6.6cm]{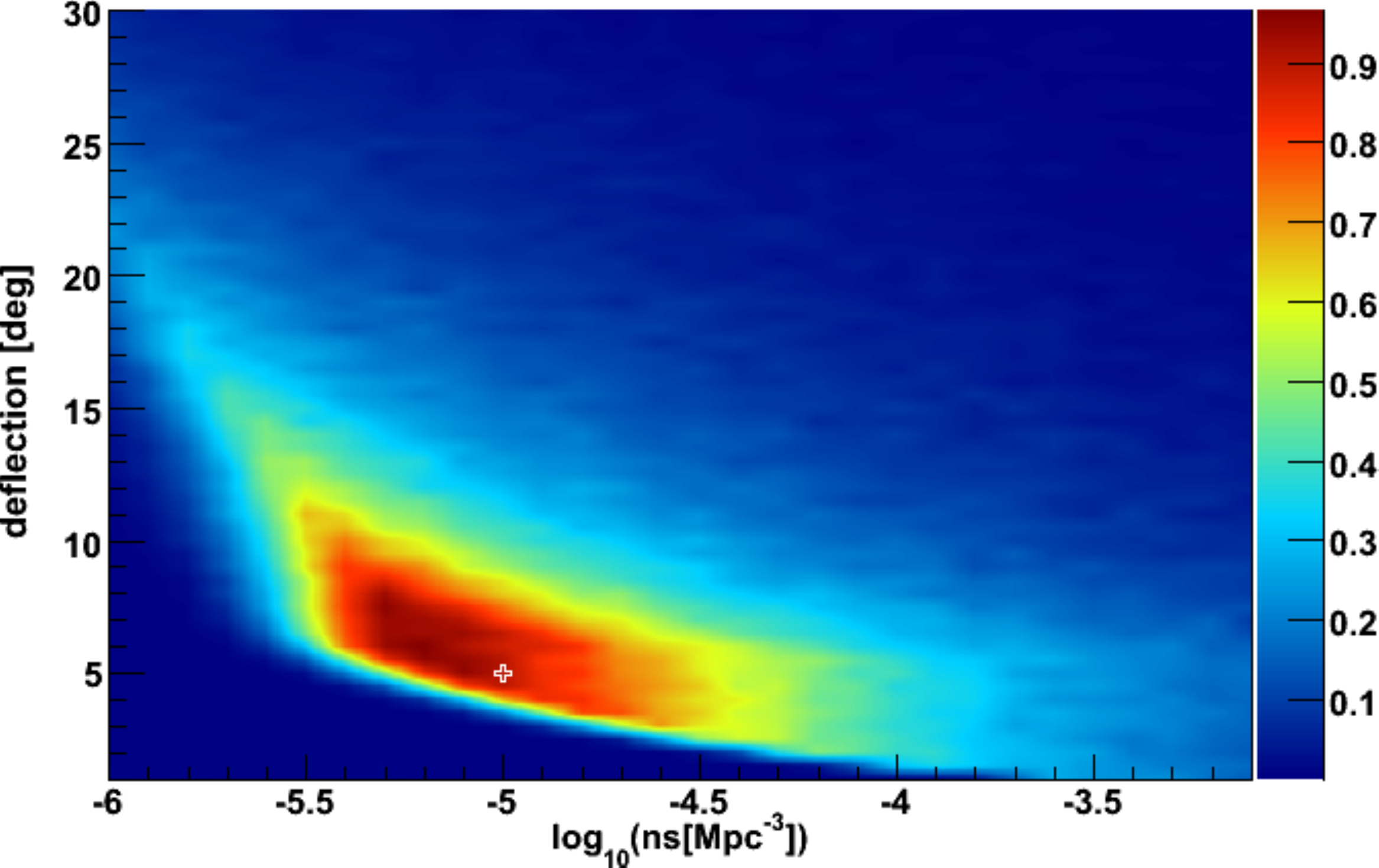}\;\;\;
\includegraphics[width=0.42\linewidth,height=6.6cm]{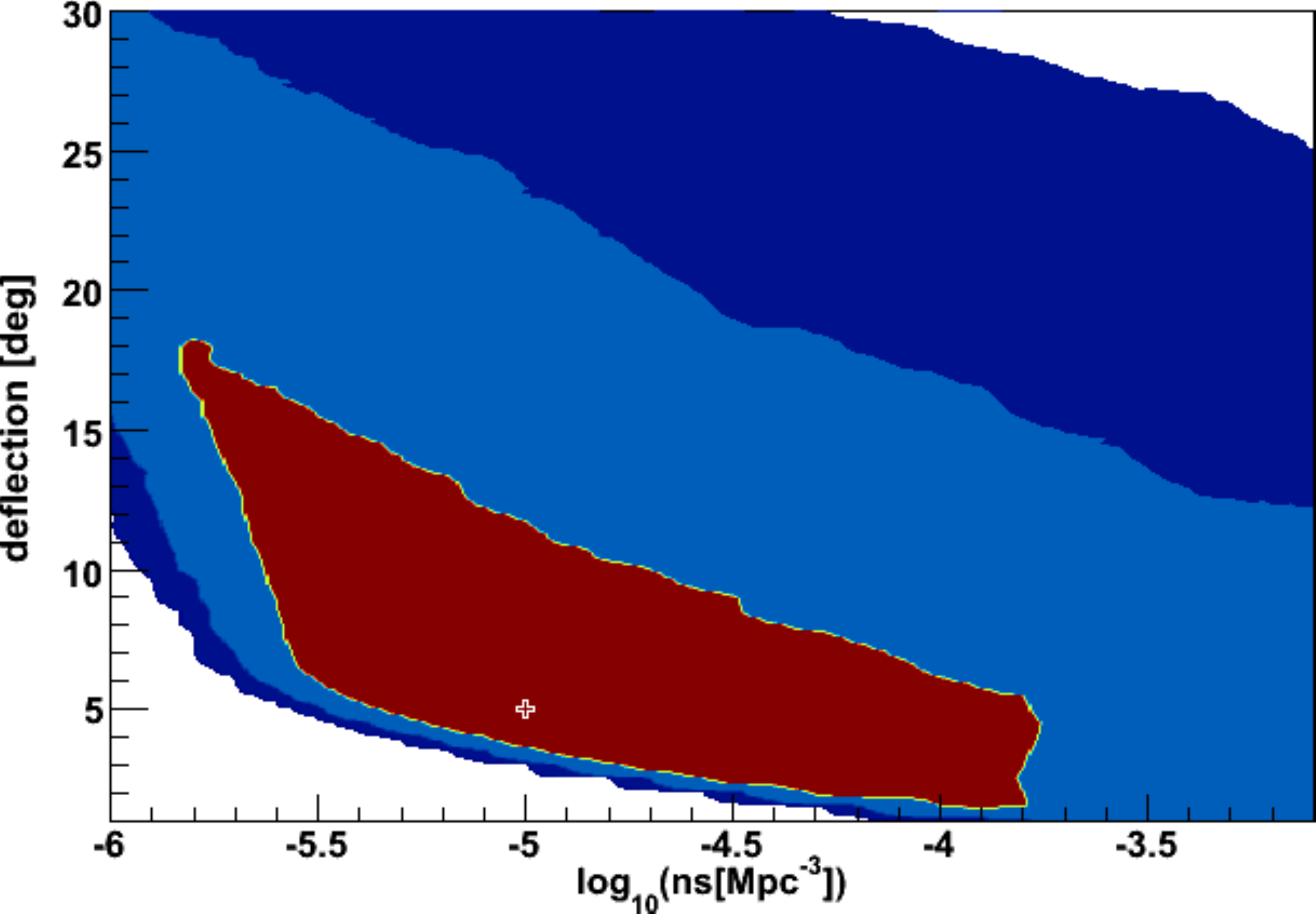}\\
\caption{Influence of the size of the data set on the astrophysical constraints derived from the clustering analysis. Left column: values of the ``clustering similarity'', $\mathcal{P}_{\pm}$, for all pairs of parameters ($n_{\mathrm{s}},\delta$), obtained from the analysis of data sets with an energy threshold $E_{\min} = 60$~EeV, containing respectively (from top to bottom) 50, 250 and 1000 events, each being a typical realization of the reference model $\mathcal{D}_{0} = (n_{\mathrm{s}} = 10^{-5}\,\mathrm{Mpc}^{-3},\delta = 5^{\circ})$, indicated by the white cross. {For all models considered here, a uniform distribution of the sources was assumed.} Right column: confidence contours extracted from the same data sets.}
\label{fig:likandconts_stats}
\end{center}
\end{figure*}

In Fig.~\ref{fig:likandconts_stats}, we show the clustering similarity ($\mathcal{P}_{\pm}$) plots, on the left, and the corresponding confidence contour plots, on the right, for data sets containing either 50, 250 or 1000 events, again built from the same model as above ($n_{\mathrm{s}} = 10^{-5}\,\mathrm{Mpc}^{-3}$, $\delta=5^{\circ}$). The two figures at the top are actually identical to those of~Fig.\ref{fig:likely_50_5_5_emin_60EeV}, but we plot them again for comparison with those obtained with larger statistics.

Not surprisingly, increasing the size of the data set allows one to draw stronger constraints and reduces the region of the parameter space that is compatible with the data, with given confidence levels. However, it can be seen that increasing the data set from 250 to 1000 events only has a moderate impact on the potential astrophysical constraints. This is because one point of the parameter space may result in very different realizations of the source position and distances. To use a language mostly used in cosmology, there is a large ``cosmic variance'' associated with the astrophysical models represented by single points in the plots, especially for low source densities. Different realizations of the same model may lead to rather dissimilar data sets, with respect to their clustering properties. Therefore, even though very large statistics will eventually give access to a very precise measurement of the actual clustering patterns associated with the UHECR sources around us, these patterns may be dissimilar to some patterns produced by other source distributions with the \emph{same} source density (and deflection angular scale), and similar to patterns produced by a number of source distributions with \emph{different} source densities.

\begin{figure}[!th]
\begin{center}
\includegraphics[width=0.87\linewidth]{Contours_50_5_5_emin_60EeV.pdf}\\
\includegraphics[width=0.87\linewidth]{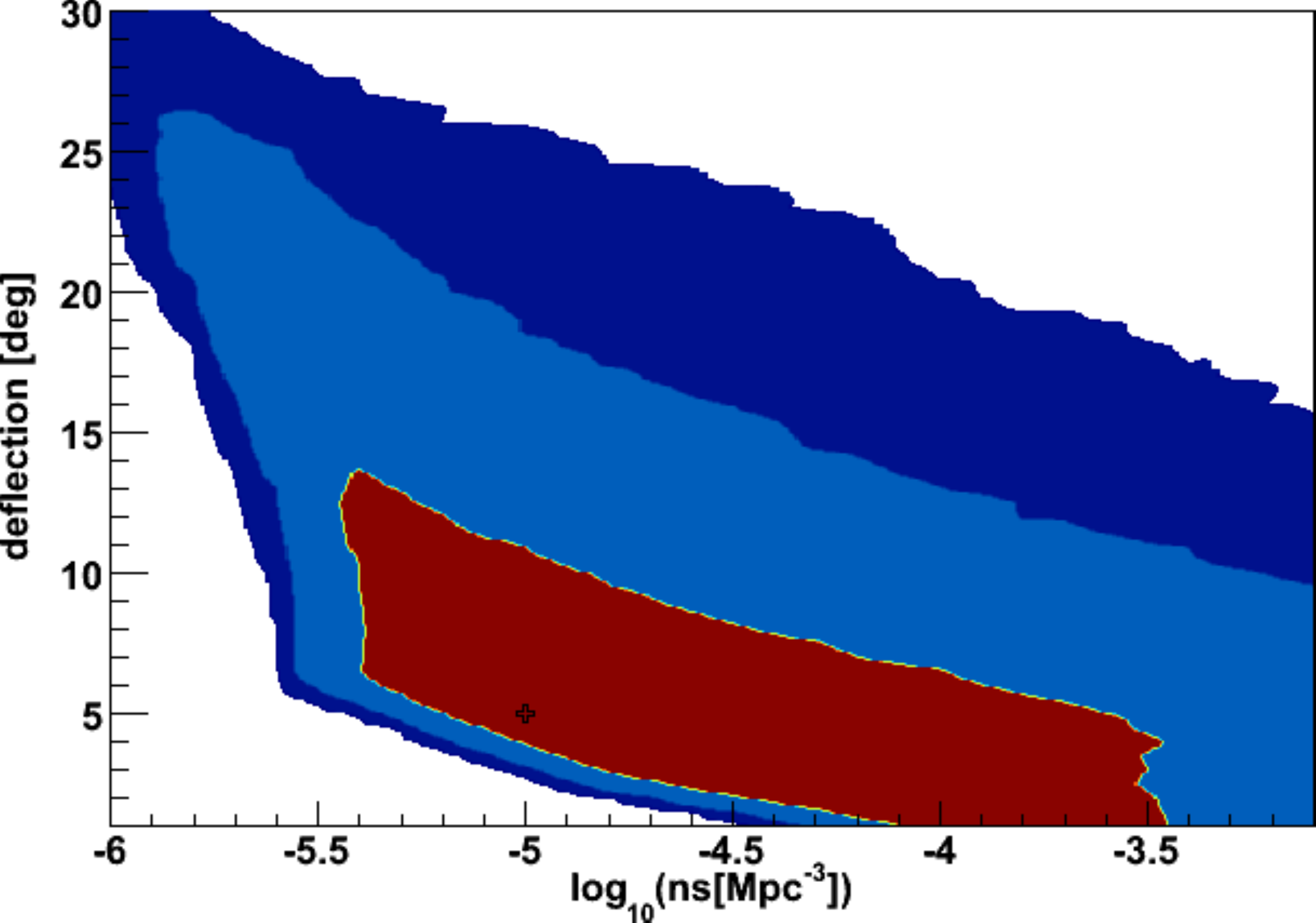}
\caption{Effect of the energy threshold $E_{\min}$ on the astrophysical constraints derived from the clustering analysis. The confidence contours are extracted from the analysis of a data set with 50 events and an energy threshold of 60~EeV (left) and 80~EeV (right), both being a typical realization of the reference model $\mathcal{D}_{0} = (n_{\mathrm{s}} = 10^{-5}\,\mathrm{Mpc}^{-3},\delta = 5^{\circ})$, indicated by the white cross. {For all models considered here, a uniform distribution was assumed.}}
\label{fig:conts_ethr}
\end{center}
\end{figure}

This illustrates a limit of the method. Since we have chosen to study the clustering properties of the data through one global parameter, $\mathcal{P}_{\pm}$, the discrimination power remains somewhat limited, even with very large statistics. On the other hand, for the same reason, the method allows one to derive interesting constraints even with limited data sets. As Fig.~\ref{fig:likandconts_stats} tentatively illustrates, this method is thus best adapted to data sets from a few tens to a few hundreds of events, which corresponds to the current and near future stage of development of UHECR observations.

\subsubsection{Energy threshold}

The size of a data set depends on the energy threshold used to build it. Whatever the total acceptance of a given experiment, the data set tends to zero as the energy increases, so there will always be an energy range where a method such as the one presented here may be relevant. As indicated above, the interest of the highest energies is that the distance from which UHECRs can reach us is more limited, so that fewer sources contribute, and characteristic clustering patterns can be explored. It is therefore interesting to investigate the potential of the method for higher energy events.

In Fig.~\ref{fig:conts_ethr}, we show the confidence contours of the clustering similarity plot for a data set of 50 events, with a minimum energy, $E_{min} = 60$~EeV (left) and 80~EeV (right), with the same model parameters as above ($n_{\mathrm{s}} = 10^{-5}\,\mathrm{Mpc}^{-3}$). Note that the former case is redrawn here from Fig.~\ref{fig:likely_50_5_5_emin_60EeV} to help comparison.

As can be seen, the size of the confidence regions is significantly smaller in the 80~EeV case. This is because the horizon scale is reduced at higher energy (typically 120~Mpc at 80~EeV vs. 190~Mpc at 60~EeV), so less sources contribute to the observed flux. Thus for a given size of the data set, one collects more events per source, which results in a larger, thus more constraining clustering signal.

\begin{figure*}[!h]
\begin{center}
\includegraphics[width=0.33\linewidth,height=5.3cm]{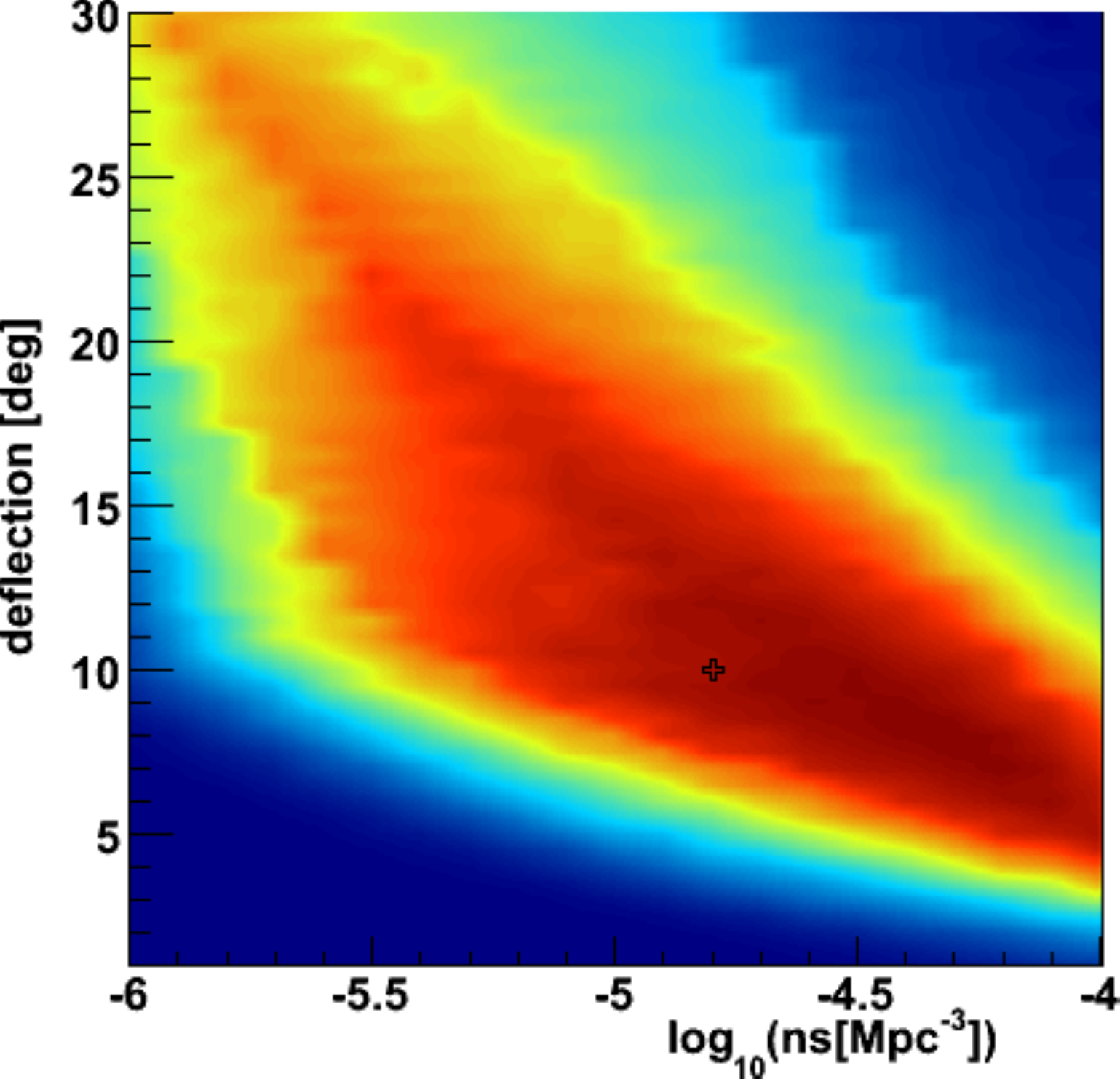}\hspace{0.9cm}
\includegraphics[width=0.33\linewidth,height=5.3cm]{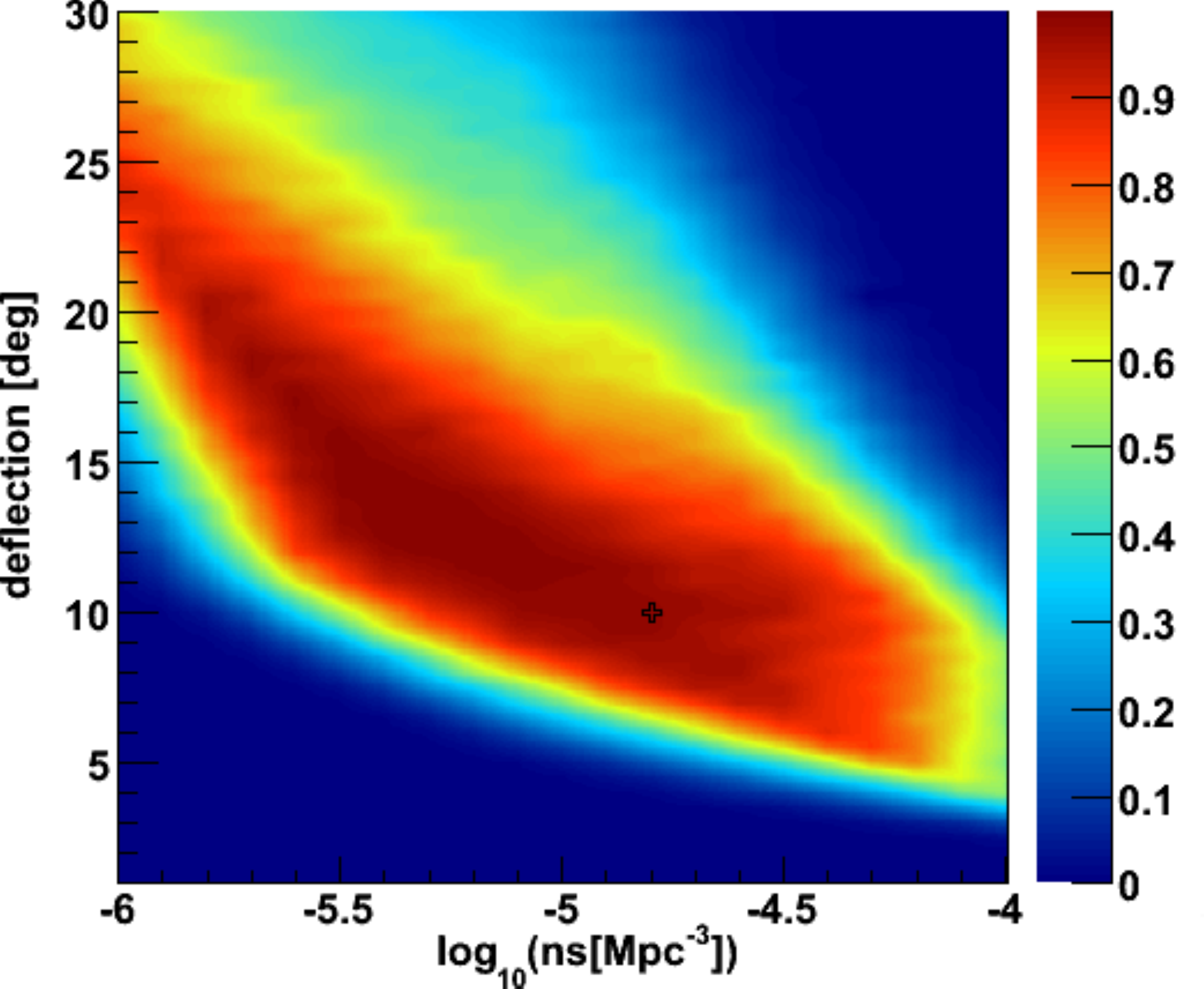}\\
\includegraphics[width=0.33\linewidth,height=5.3cm]{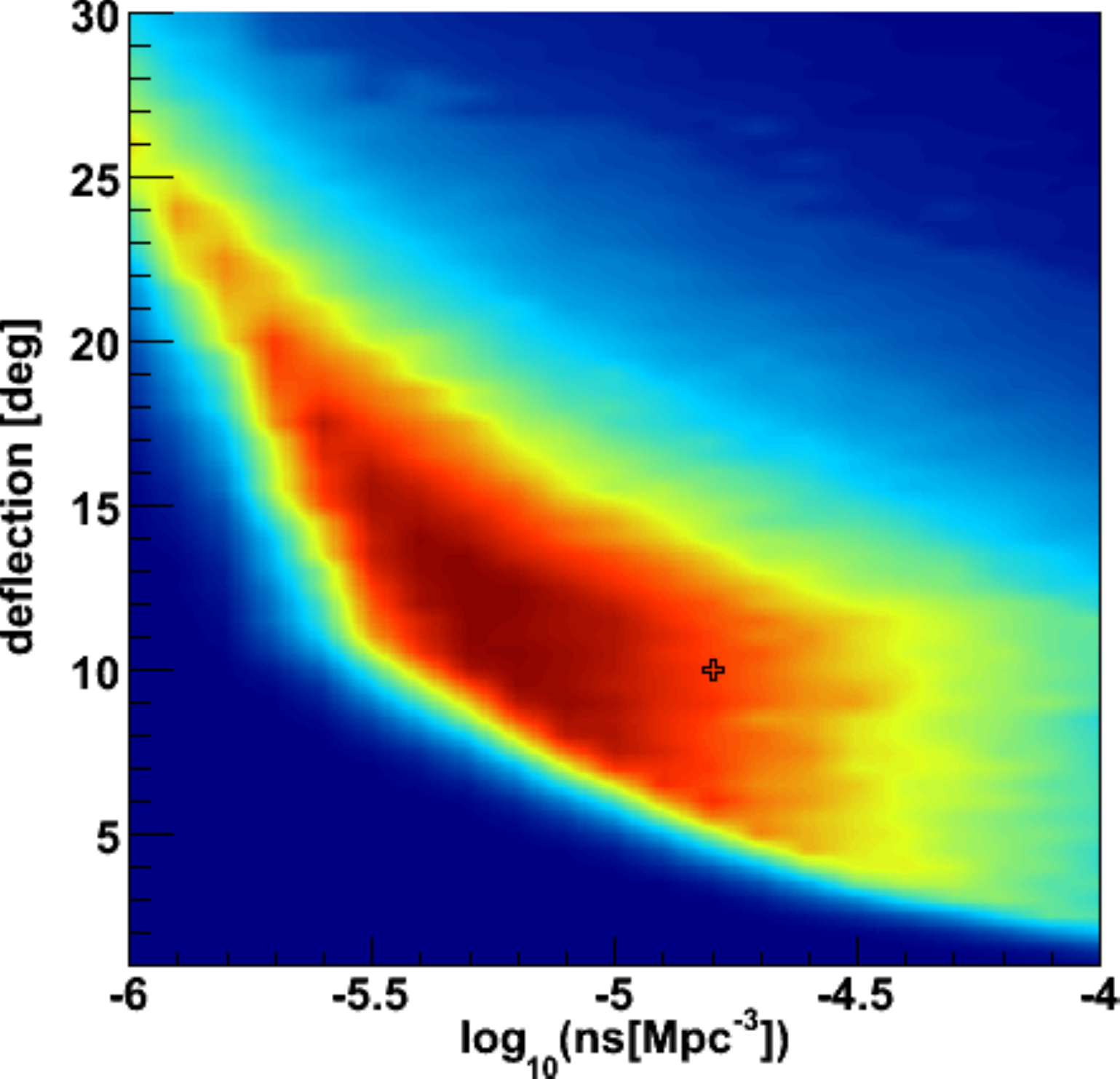}\hspace{0.9cm}
\includegraphics[width=0.33\linewidth,height=5.3cm]{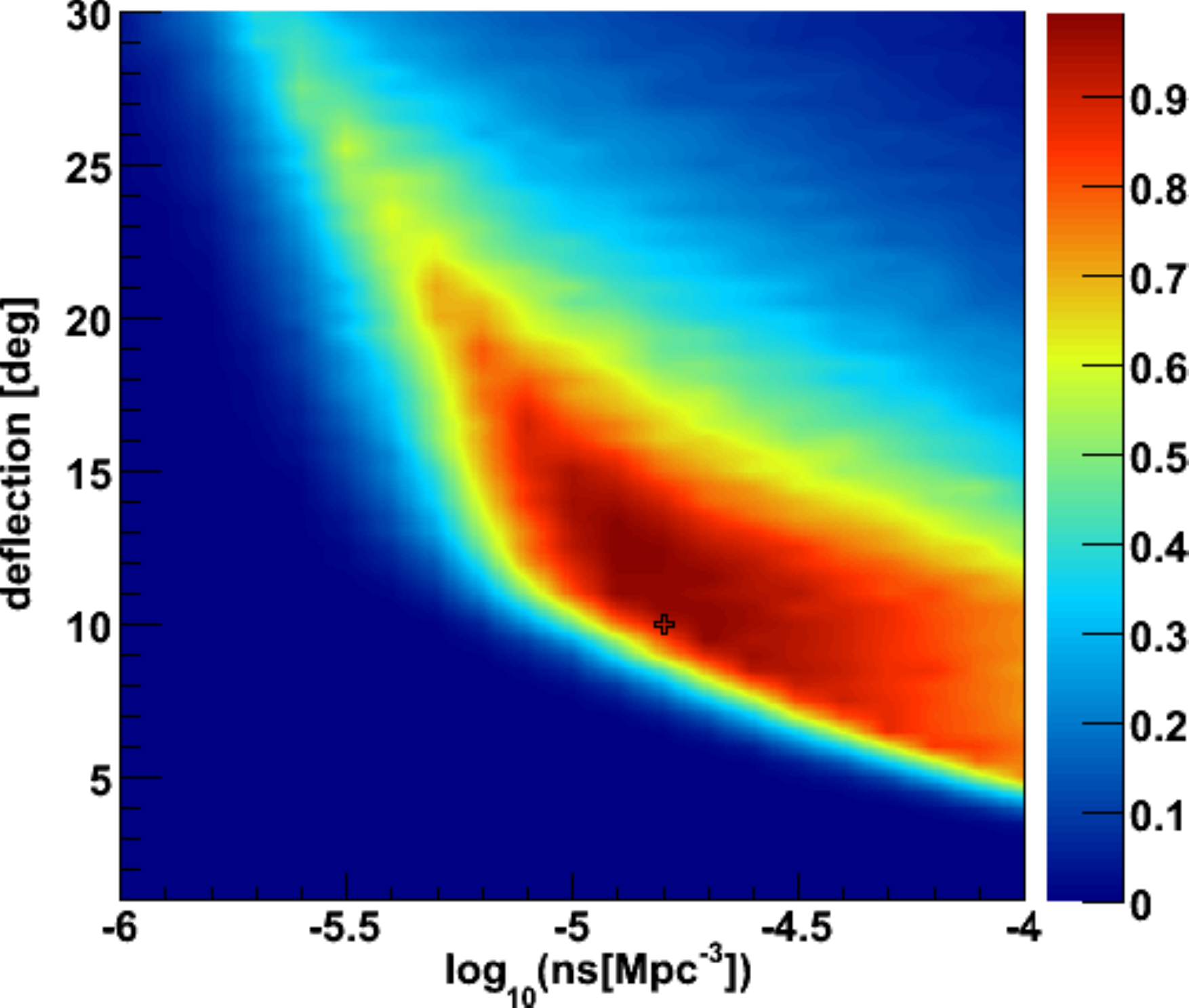}\\
\includegraphics[width=0.33\linewidth,height=5.3cm]{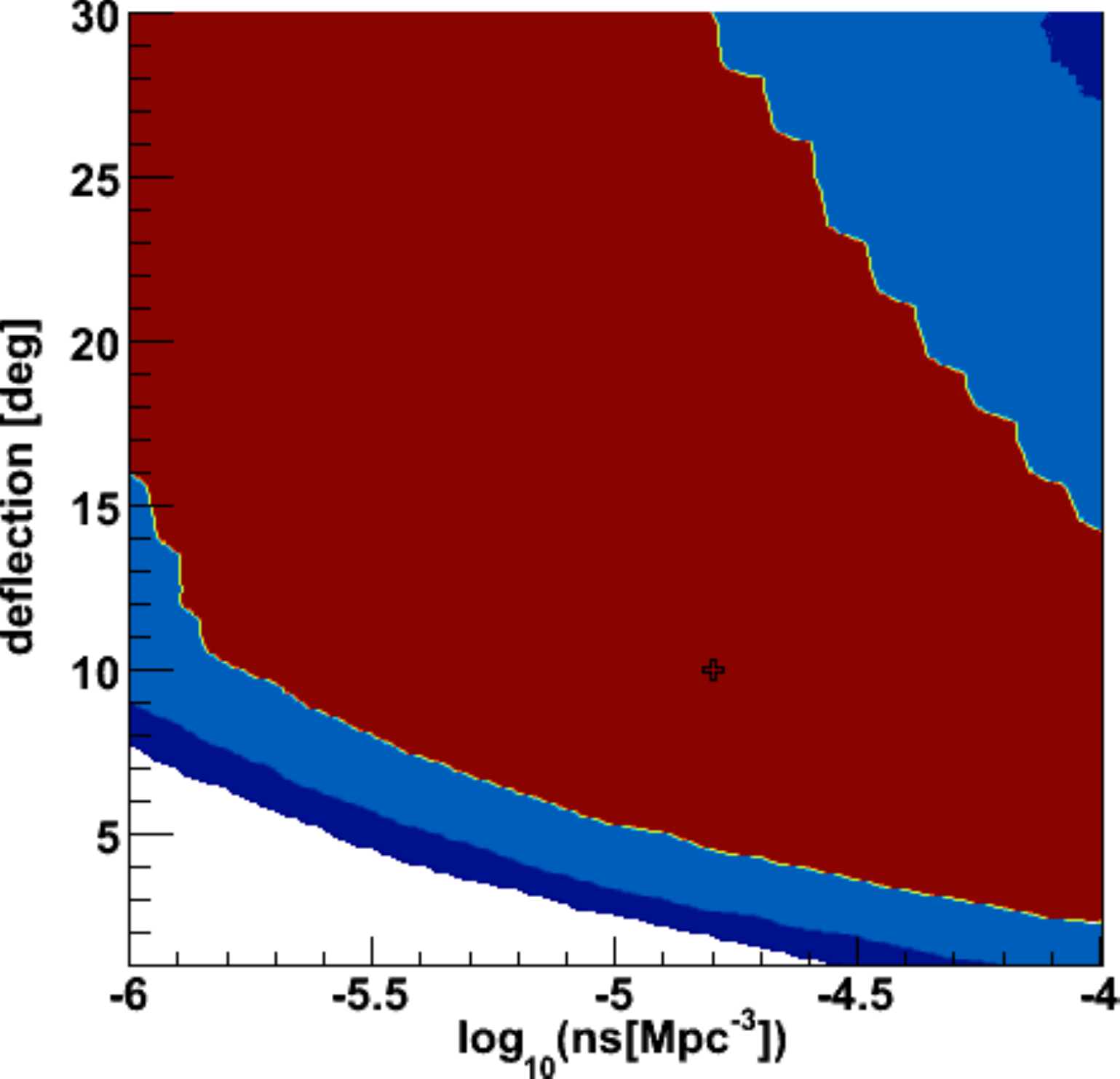}\hspace{0.8cm}
\includegraphics[width=0.33\linewidth,height=5.3cm]{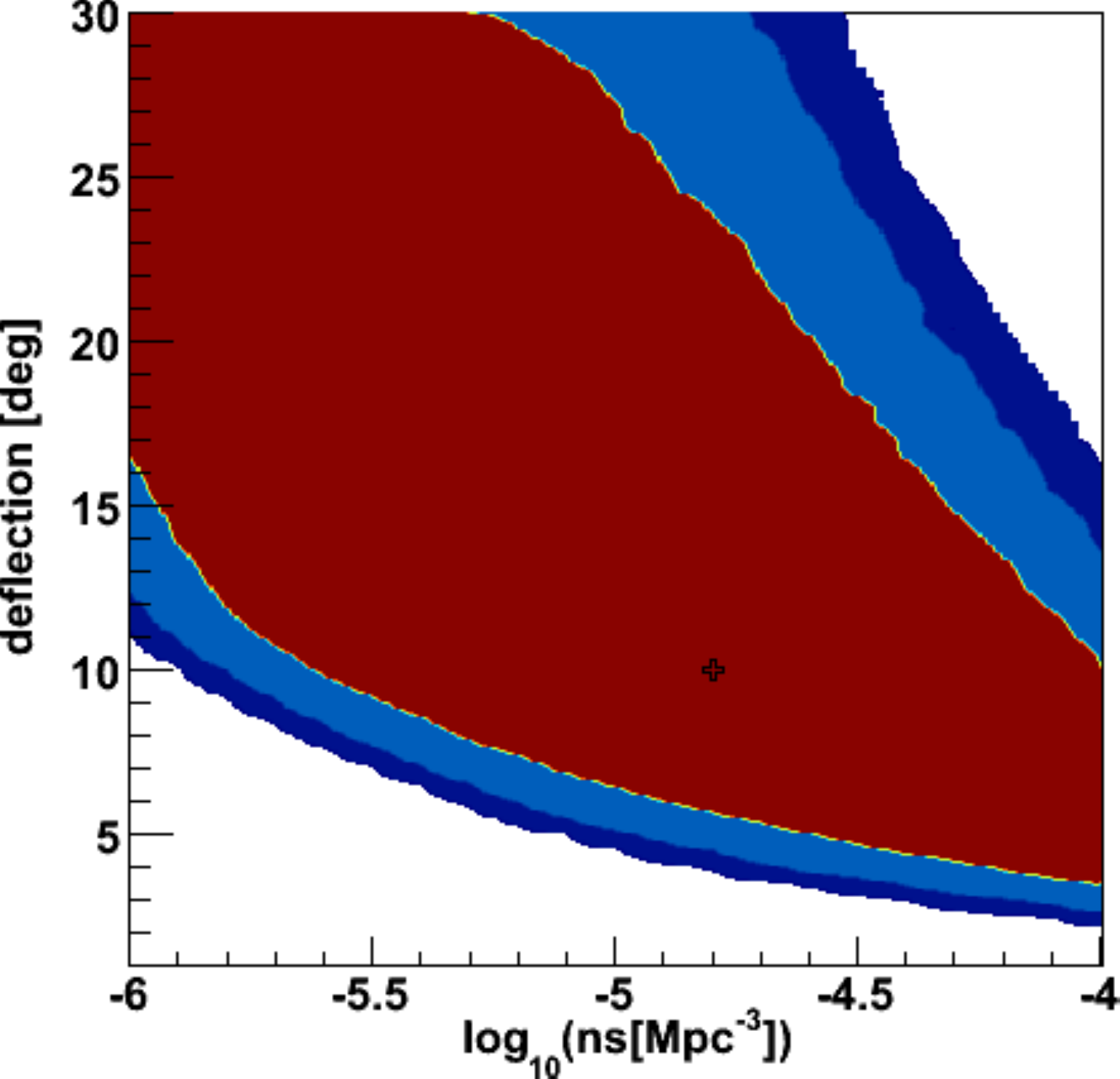}\\
\includegraphics[width=0.33\linewidth,height=5.3cm]{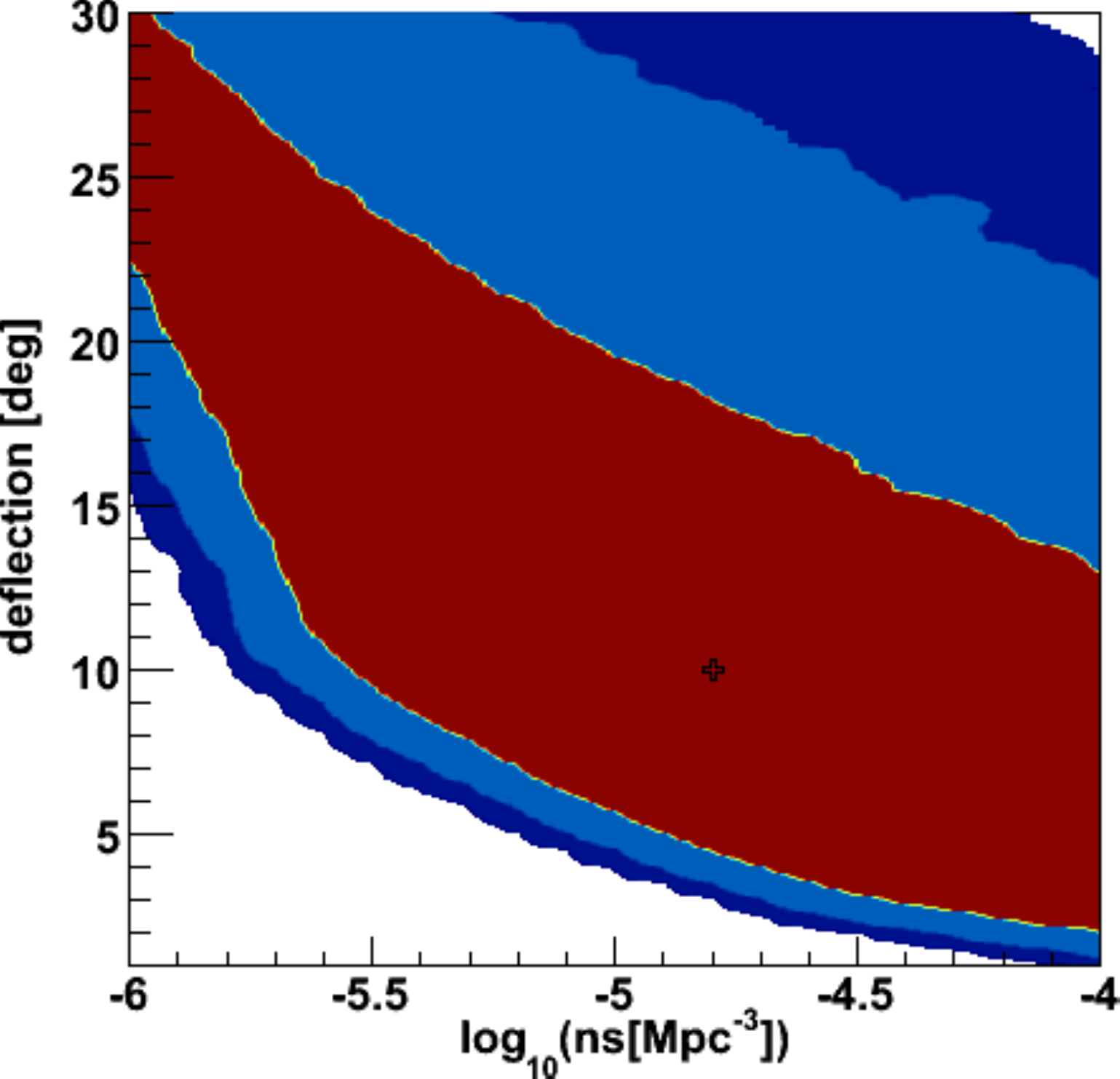}\hspace{0.8cm}
\includegraphics[width=0.33\linewidth,height=5.3cm]{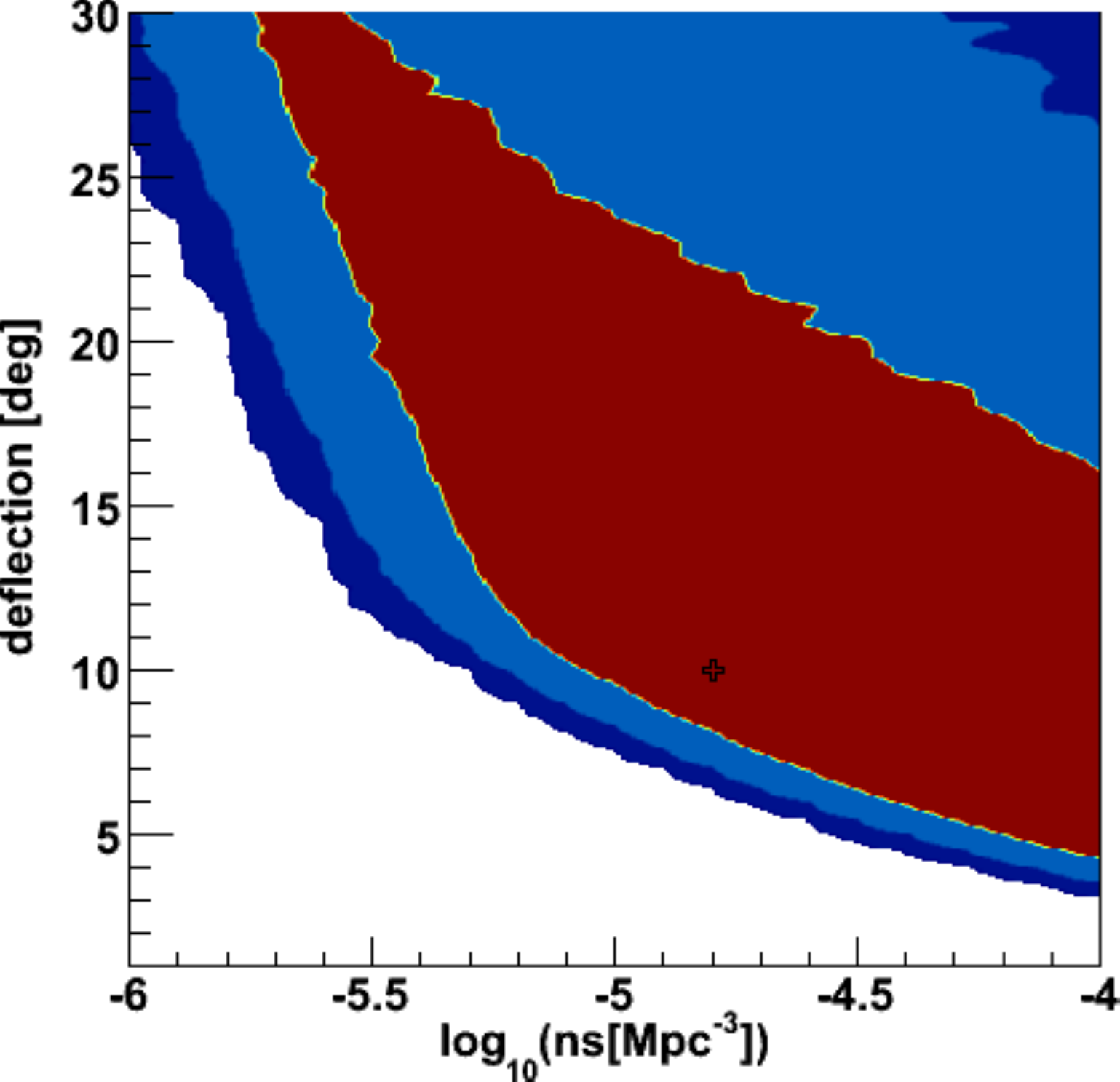}
\caption{Values of ``clustering similarity'', $\mathcal{P}_{\pm}$, (top four figures) and confidence contours (bottom four figures) for all ($n_{\mathrm{s}},\delta$) obtained with the analysis of 4 data sets with different assumptions, each is a typical realization of the reference model $\mathcal{D}_{0}$=($n_{\mathrm{s}} = 10^{-4.8}\,\mathrm{Mpc}^{-3}$,$\delta = 10^{\circ}$) indicated by a black cross. Rightwards and downwards, the number of events and energy threshold in EeV are: $(N,E_{\min}) = (50,60); (250,60); (50,80); (250,80)$. The source distribution follows the galaxies and the acceptance mimics the coverage of the Pierre Auger Observatory.}
\label{fig:another_example}
\end{center}
\end{figure*}

Of course, the higher the minimum energy considered the larger the exposure required to achieve the same size of the data set. In practice, there is thus a trade off between the constraining power gained by concentrating on the highest energies and that gained by increasing the size of the data set. Here, we do not address this specific question in detail. Nevertheless, for the sake of completeness, we show in Fig.\ref{fig:another_example} the results of the analysis corresponding to four additional cases: 50~events above 60~EeV, 50~events above 80~EeV, 250~events above 60~EeV, and 250~events above 80~EeV. The reference model was chosen to have $n_{\mathrm{s}} = 10^{-4.8}\,\mathrm{Mpc}^{-3}$ and $\delta=10^{\circ}$ (as indicated by the cross on the plots), with an underlying source distribution following that of the 2MRS galaxy catalog.

\section{Conclusion}

In this paper, we have explored how the usual clustering analysis of a given UHECR data set could be extended to constrain the responsible astrophysical models in a 2D parameter space consisting of the effective density of UHECR sources and the effective angular deflection of the UHECRs from their source to the Earth.

We first studied the sensitivity of a typical UHECR detector to anisotropies and showed that detectors with a moderate acceptance can detect a significant anisotropy signal in a large number of astrophysical cases, or, failing that, constrain a large class of astrophysical models.

We then introduced an indicator, loosely referred to as the ``clustering similarity'' of a given data set with respect to sample data sets built from a given class of astrophysical models. This allowed us to propose an improved method to make the most of the clustering analysis based on the two-point correlation function.

We applied this method to artificial data sets, demonstrating its statistical power and studying the influence of parameters such as the size of the data set, the minimum energy considered and the spatial distribution of the UHECR sources.

Since our method condenses all the information contained in the two-point correlation function into one estimator, it can be powerful even with small data sets, corresponding to the current experimental situation in the GZK energy range. For larger statistics, more refined studies (e.g. searching for energy-ordered multiplets) will be complementary and help gather more information about the UHECR sources in the universe.

Finally, let us note that the whole study assumes that one can identify one typical deflection angular scale and one typical source density, in other words, that the parameter space that we explore is indeed relevant to the astrophysical situation under investigation. This may not be the case for UHECRs, either because of a distribution of sources with a luminosity function following a power law, thus with no clear source density, or because there are more than one component with different typical scales. In that case, however, applying this method to actual UHECR data sets still allows one to constrain the phenomenological models corresponding to the regions of the parameter space that would be in conflict with the clustering analysis.

{Note also that in the case of transient sources (such as gamma-ray burst or AGN flares,  ), what we call here the source density is but the density of sources that can actually contribute to the UHECR flux observed at the current time. This effective density depends on the occurrence (or repetition) rate of the transient events, $1/\tau_{\mathrm{occ}}$, compared to the time spread, $\Delta t_{\mathrm{obs}}$, of the expected signal (see e.g. Murase \& Takami, 2009 for the implications of UHECRs data on such transient sources). The latter happens to be related to the deflection scale, but in any case, the effective (instantaneous) density of sources can be obtained as $n_{\mathrm{s,eff}} = n_{\mathrm{s}}\times \Delta t_{\mathrm{obs}}/\tau_{\mathrm{occ}}$. This is the source density that should appear in the $x$-axis of our contours plots.}

\end{document}